\documentclass[structabstract]{aa}  
\usepackage{graphicx}
\usepackage{amssymb}
\usepackage{amsmath}
\usepackage[english]{babel}
\usepackage[latin1]{inputenc}
\usepackage{latexsym}
\usepackage{verbatim}
\usepackage[colorlinks=true,dvips]{hyperref}
\usepackage{lscape}

\begin{document}
   \title{Accretion disks around black holes in modified strong gravity}


   \author{Daniela P\'erez
          \inst{1}
          \and
          Gustavo E. Romero{\inst{1}\: \inst{2}}
          \and Santiago E. Perez Bergliaffa\inst{3}}

   \institute{Instituto Argentino de Radioastronomía,Camino Gral Belgrano Km 40\\
              C.C.5, (1984) Villa Elisa, Bs. As., Argentina\\
              \email{danielaperez@iar.unlp.edu.ar}
         \and
             Facultad de Ciencias Astron\'omicas y Geof{\'\i}sicas, UNLP, Paseo del Bosque s/n\\
             CP (1900), La Plata, Bs. As., Argentina\\
             \email{romero@iar.unlp.edu.ar}
          \and
          Departamento de Física Teórica, Instituto de Física, Universidade do Estado do Rio de Janeiro,\\
 					Rua São Francisco Xavier 524, Maracanã Rio de Janeiro - RJ, Brasil, CEP: 20550-900\\
 					\email{sepbergliaffa@gmail.com}}

   \date{}

 
  \abstract{Stellar-mass black holes offer what is perhaps the best scenario to test theories of gravity in the strong-field regime. In particular, $f(R)$ theories, which have been widely discus
   in a cosmological context, can be constrained through realistic astrophysical models of phenomena around black holes.}{We aim at building radiative models of thin accretion disks for both Schwarzschild and Kerr black holes in $f(R)$ gravity.}{We study particle motion in 
$f(R)$-Schwarzschild and Kerr space-times.}
{We present the spectral energy distribution of the accretion disk around constant Ricci scalar $f(R)$ black holes, and constrain specific $f(R)$ prescriptions using features of these systems.}{A precise determination of both the spin and accretion rate onto black holes along with X-ray observations of their thermal spectrum might allow to identify deviations of gravity from General Relativity. We use recent data on the high-mass X-ray binary Cygnus X-1 to restrict the values of the parameters of a class of $f(R)$ models.}


\titlerunning{Accretion disks onto black holes}
\authorrunning{Daniela P\'erez et al.}

\maketitle

\section{Introduction}


General Relativity (GR) is consistent, in some cases with good precision, with observational results (see for instance Will 2006). However, an energy momentum tensor representing 
exotic matter (loosely called `dark energy', e.g. Li, Li, and Wang 2011) must be introduced in the right hand side of Einstein's equations
to fit the currently available data when these are interpreted in the framework of the standard cosmological model (based on GR).
Dark energy can be
modelled by a cosmological constant, or by a scalar field
with an equation of state given by $p = \omega_{\rm DE} \rho$, 
where 
$\omega_{\rm DE} < -1/3$ (Biswas et al. 2010a, 2010b). None of these descriptions is free of problems, since the energy density associated with the cosmological constant that is inferred from astronomical observations is approximately 120 orders of magnitude lower than the value predicted by field theory (e.g. Weinberg 1989, Capozziello and Faraoni 2010), whereas the scalar field has features that are at odds with  
the scalar fields of particle physics (Sotiriou and Faraoni 2010).

A different approach to explain the cosmological data is to modify the field equations of the gravitational field,
in such a way that 
the ensuing theory differs from Einstein's in the low-curvature regime.
Since there is no \textsl{a priori} fundamental reason to restrict the gravitational Lagrangian to a linear function of the Ricci scalar $R$ (see for instance Magnano et al. 1987),  
more general theories can be formulated using nonlinear functions of this scalar. The so-called $f(R)$ theories (e.g. Capozziello and Faraoni 2010) were first used to mimic the inferred accelerated expansion of the universe by 
Capoziello (2002). Currently there is in the literature a handful of $f(R)$-models in agreement with available data (De Felice and Tsujikawa 2010). 

Although the present 
revival of $f(R)$ theories is mainly due to their use in the description 
of phenomena
that 
take place for low values of the Riemann curvature,
these theories have also been applied to gravity 
in the opposite regime. As there is no direct evidence of the behaviour of the gravitational field for very large values of the curvature, 
the early universe and compact objects offer the possibility
to find deviations from GR.
Among the studies in modified gravity in the strong regime, we can mention the successful inflationary model based on the $R+\alpha R^2$ theory 
(Starobinsky 1980), and the related studies of reheating 
(Motohashi and Nishizawa 2012)
and particle production 
(Arbuzova, Dolgov, and Reverberi 2012) in the early universe. 
Also of importance is the treatment of 
neutron stars (Cooney, DeDeo and Psaltis, 2010) and
black hole solutions.

Different aspects of black hole physics in $f(R)$ theories have been discussed in the literature by Psaltis et al. (2008), 
Hendi \& Momeni (2011), 
Myung (2011), Myung et al. (2011), 
Moon, Myung, and Son (2011a, b), Hendi et al. (2012), and
Habib Mazharimousavi et al. (2012).
Static and spherically symmetric black hole solutions were obtained via perturbation theory by 
de la Cruz-Dombriz, Dobado, and Maroto (2009), 
whereas black holes with these symmetries have been studied by means of a near-horizon analysis by Perez Bergliaffa \& De Oliveira (2011). 
Finally, 
$f(R)$-Kerr-Newman black holes solutions with constant Ricci scalar 
have been recently studied by Cembranos and collaborators (Cembranos et al. 2011). 

From an astrophysical point of view,
thin accretion
for the Schwarszchild space-time in $f(R)$-gravity has been discussed by 
Pun et al. (2008), without including the expected spectra of concrete astrophysical black holes or any comparison with observational data.  

In the present work we investigate the existence of stable circular orbits in Schwarzschild and Kerr $f(R)$ space-times with constant Ricci scalar and analyze the main features of accretion disks around these black holes. In particular, we present temperature and spectral energy distributions
for Page-Thorne disks, and compare the results with those obtained using the standard Shakura-Sunyaev model.

The paper is organized as follows. In Section
\ref{bhs} we provide a brief review of $f(R)$ theories of gravity. Circular orbits in both Schwarzschild and Kerr $f(R)$ space-times with constant Ricci scalar are studied in Section  
\ref{orbits}. Section \ref{sacc} is devoted to the calculation of the properties of accretion disks in these space-times. Consequences 
for some specific prescriptions for the function $f$
are discussed in Section \ref{spec}. We close with some considerations
on the potential of astronomical observations to test the strong-field regime of gravity.

\section{$f(R)$ gravity}
\label{bhs}
In $f(R)$ gravity, the Lagrangian of the Hilbert-Einstein action, given by:
\begin{equation}
S[g] = \frac{c^{3}}{16\pi G}\int  R \sqrt{-g}\; d^{4}x,
\end{equation}
is generalized to:
\begin{equation}\label{lagrafr}
S[g] = \frac{c^{3}}{16\pi G}\int \left(R +f(R)\right) \sqrt{-g}\; d^{4}x,
\end{equation}
where $g$ is the determinant of the metric tensor, and $f(R)$ is an arbitrary function of the Ricci scalar. In the metric formalism the field equations are obtained varying Eq. (\ref{lagrafr}) with respect to the metric:
\begin{eqnarray}\label{fr-eqs}
 & &R_{\rm \mu \nu}(1+f'(R)) - \frac{1}{2}g_{\rm \mu \nu} (R+f(R))\\ \nonumber
 &+ &(\nabla_{\rm \mu}\nabla_{\rm \nu}-g_{\rm \mu \nu}\Box)f'(R) + \frac{16\pi G}{c^{4}}\;\; T_{\rm \mu \nu}=0,
\end{eqnarray}
where $R_{\rm \mu \nu}$ is the Ricci Tensor, $\Box \equiv \nabla_{\rm \beta} \nabla^{\rm \beta}$, $f'(R) = df(R)/dR$, and the energy momentum tensor is defined by:
\begin{equation}
T_{\rm \mu \nu} = \frac{-2}{\sqrt{-g}}\frac{\delta(\sqrt{-g}{\cal L}_{\rm m})}{\delta g^{\mu \nu}}.
\end{equation}
Here, ${\cal L}_{\rm m}$ stands for the matter Lagrangian.

Equations (\ref{fr-eqs}) are a system of fourth-order nonlinear equations for the metric tensor $g_{\rm \mu \nu}$. An important difference between them and the Einstein field equations is that in $f(R)$ theories the Ricci scalar $R$ and the trace $T$ of the energy momentum tensor are differentially linked, as can be seen by taking the trace of Eq. (\ref{fr-eqs}), which yields:
\begin{equation}
R (1+f'(R))-2(R+f(R))-3\Box f'(R)+ \frac{16\pi G}{c^{4}} \;T =0.
\label{trace}
\end{equation}
Hence, depending on the form of the function $f$, there may be solutions with traceless energy-momentum tensor and nonzero Ricci scalar. This is precisely the case of black hole space-times in the absence of a matter source. 

Notice that in the case of constant Ricci scalar $R_0$ without matter sources, Eqs. (\ref{fr-eqs}) can be re-written as: 
$$
R_{\mu\nu} = \Lambda g_{\mu\nu},
$$
where: 
\begin{equation}
\Lambda \equiv \frac{f(R_0)}{f'(R_0)-1},
\label{coco}
\end{equation}
and, by Eq.(\ref{trace}):
\begin{equation}
R_0=\frac{2f(R_0)}{f'(R_0)-1}.
\label{condi}
\end{equation}
Hence, in this case, any $f(R)$ theory is formally, but not physically, equivalent to GR with a cosmological constant given by Eq.(\ref{coco}) \footnote{Accretion through thick disks onto Schwarzschild and Kerr black holes with a repulsive cosmological constant was studied by Rezzolla, Zanotti and Font (2003), and Slan\'y and Stuchl\'{\i}k (2005), respectively. These studies do not 
analyze spectra or compare results with observational data.}.

We shall investigate the existence of stable circular orbits in Schwarzschild and Kerr $f(R)$ space-times with constant Ricci scalar in the next section. 


\section{Circular orbits around a black hole in $f(R)$}
\label{orbits}

\subsection{$f(R)$-Schwarzschild space-time}

The Schwarzschild space-time metric in $f(R)$ theories with constant Ricci scalar $R_{{0}}$ takes the form (Cembranos et al. 2011):
\begin{eqnarray}\label{g-Sc}
ds^{2}& = &-\left[\left(c^{2}-\frac{2GM}{r}\right)-\frac{c^{2}R_{{0}}}{12}r^{2}\right]\\ \nonumber dt^{2}
& + & \frac{dr^{2}}{\left[\left(1-\frac{2MG}{c^{2}r}\right)-\frac{R_{{0}}}{12}r^{2}\right]} 
 +  r^{2} \left(d\theta^{2}+\sin^{2}\theta d\phi^{2}\right),
\end{eqnarray}
where $R_{{0}}$, given by Eq. (\ref{condi}), can take, in principle, any real value. Since we are looking for space-time metrics that may represent astrophysical black holes, we shall select those values of $R_{{0}}$ that lead to acceptable solutions.

The radius $r_0$ of the horizon follows from the condition $g_{00}(r_0)=0$. From Eq. (\ref{g-Sc}), the values of $r$ corresponding to the event horizon satisfy:
\begin{equation}\label{hori}
c^{2}R_{0} r^{3}-12c^{2} r+24GM = 0.
\end{equation}
In terms of the following adimensional quantities:
\begin{eqnarray}
\mathsf{x} & \equiv & \frac{r}{r_{{\rm g}}},\\
\mathsf{R_0} & \equiv & R_{{0}}r_{{\rm g}}^{2},
\end{eqnarray}
where $r_{{\rm g}} = GM/c^2$, this equation  
takes the form:
\begin{equation}
\mathsf{R_0}\mathsf{x}^{3}-12\mathsf{x}+24 = 0.
\end{equation}


\begin{figure}
\center
\includegraphics[width=8cm]{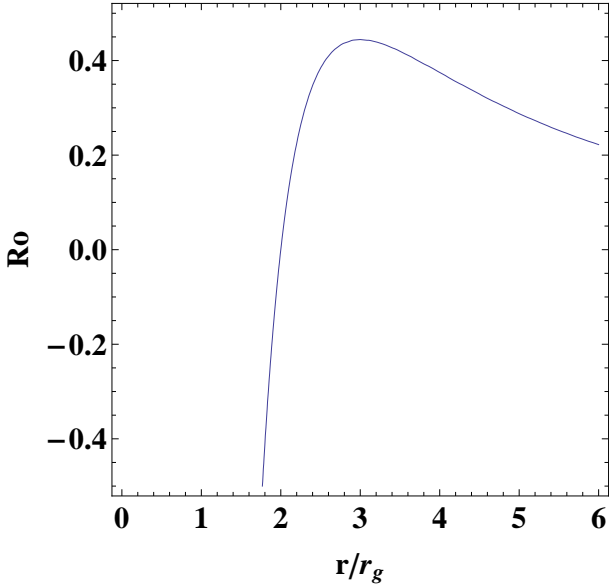}
\caption{Plot of the Ricci scalar as a function of the radial coordinate of the event horizon in $f(R)$-Schwarzschild space-time.}
\label{hori1}
\end{figure}

We show in Figure \ref{hori1} the Ricci scalar as a function of the radial coordinate of the event horizon. We see that for $\mathsf{R_0} \in (0,4/9)$ there is an inner black hole event horizon and an outer cosmological horizon, whereas for $\mathsf R_0 \leq 0$ there is only one black hole event horizon. The event and cosmological horizons collapse for $\mathsf R_0 = 4/9$ and for larger values of the Ricci scalar naked singularities occur and hence, no black holes are possible\footnote{The results obtained in our analysis of $f(R)$-Schwarzschild space-time with constant Ricci scalar are consistent with those given by Stuchl{\'\i}k and Hled{\'\i}k (1999) in Schwarzschild-de Sitter and Schwarzschild-anti de Sitter space-times.}. We shall, then, restrict ourselves to the study of trajectories for $\mathsf{R_0} \in (-\infty,4/9]$.

\subsubsection{Equations of motion and effective potential in $f(R)$-Schwarzschild space-time}  

The geodesic equations in the metric given in Eq. (\ref{g-Sc}) can be obtained by means of the Euler-Lagrange equations using the Lagrangian:
\begin{eqnarray}\label{lagr-sc}
L & = & - \left[\left(c^{2}-\frac{2GM}{r}\right)-\frac{c^{2}R_{{0}}}{12}r^{2}\right] \dot{t}^{2}\\ \nonumber
 & + & \frac{1}{\left[\left(1-\frac{2GM}{c^{2}r}\right)-\frac{R_{{0}}}{12}r^{2}\right]} \dot{r}^{2}
+r^{2}\left(\dot{\theta}^{2}+{\sin \theta}^{2} \dot{\phi}^{2}\right),
\end{eqnarray}
where $\dot{x}^{\mu} \equiv dx^{\mu}/d\sigma$, and $\sigma$ is an affine parameter along the geodesic $x^{\mu}(\sigma)$.
The resulting geodesic equations for $t$ and $\phi$ are (setting 
$\theta = \pi/2$):
\begin{eqnarray}
\left[\left(1-\frac{2GM}{c^{2}r}\right)-\frac{R_{{0}}}{12}r^{2}\right] \dot{t} & = & k,\label{et}\\
r^{2} \dot{\phi} = h\label{ephi},
\end{eqnarray}
where $k$ and $h$ are constants. An equation for $r$ that is simpler than the one obtained from the Lagrangian follows from the modulus of the 
4-momentum $\textbf{p}$, given by $
g_{\mu\nu} x^{\mu} x^{\nu} = \epsilon^{2}$,
where $\epsilon^{2}=c^{2}$ for massive particles, and $\epsilon^{2}=0$ for photons. It takes the form:
\begin{eqnarray}\label{er}
& - &\left[\left(c^{2}-\frac{2GM}{r}\right)-\frac{c^{2}R_{{0}}}{12}r^{2}\right] \dot{t}^{2}\\ \nonumber
& + &\frac{\dot{r}^{2}}{\left[\left(1-\frac{2GM}{c^{2}r}\right)-\frac{R_{{0}}}{12}r^{2}\right]} + r^{2} \dot{\phi}^{2} = \epsilon^{2},
\end{eqnarray}
with $\dot t$ and $\dot \phi$ given by 
Eqs. (\ref{et}) and (\ref{ephi}), respectively.
The set of Eqs. (\ref{et}), (\ref{ephi}), and (\ref{er}) completely determine the motion of a particle in the $f(R)$-Schwarzschild space-time.

\subsubsection{Trajectories of massive particles}\label{orbits-sch}

Equations (\ref{et}), (\ref{ephi}), and (\ref{er})
can be used to obtain
the so-called energy equation (e.g. Hobson et al. 2006):
\begin{eqnarray}\label{energy1}
& &\dot{r}^{2}+\frac{h^{2}}{r^{2}} \left[\left(1-\frac{2GM}{c^{2}r}\right)-\frac{R_{{0}}}{12}r^{2}\right]\\ \nonumber 
&+&\left(-\frac{2GM}{r}-\frac{c^2R_{{0}}}{12}r^{2}\right) = c^{2}(k^{2}-1).
\label{energy}
\end{eqnarray}
The constant $k$ is defined as $ k= E/m_{0}c^{2}$, where $E$ represents the total energy of the particle in its orbit, and $m_{0}c^{2}$ its rest mass energy. The constant $h$ stands for the angular momentum of the particle per unit mass.
From Eq. (\ref{energy1}) we can identify the effective potential per unit mass as:
\begin{eqnarray}\label{pot}
V_{{\rm eff}}(r)  & = & \frac{h^{2}}{2 r^{2}} \left(1-\frac{2GM}{c^{2}r}-\frac{R_{{0}}}{12}r^{2}\right) +\\ \nonumber
& & \frac{1}{2}\left(-\frac{2GM}{r}-\frac{c^2R_{{0}}}{12}r^{2}\right).
\end{eqnarray}
The extrema of the effective potential are obtained by looking for the roots of the derivative of the latter equation with respect to the radial coordinate.
In terms of $\mathsf{x}$, $\mathsf{R_{{0}}}$, and
the adimensional angular momentum per unit mass of the particle 
$\mathsf{h}=
h(cr_{{\rm g}})^{-1}$,
this reads:
\begin{equation}\label{vf1}
\frac{dV_{{\rm eff}}}{d\mathsf{x}} =c^{2}\left( - \frac{\mathsf{h}^{2}}{\mathsf{x}^{3}}+\frac{3\mathsf{h}^{2}}{\mathsf{x}^{4}}+\frac{1}{\mathsf{x}^{2}}
-\frac{\mathsf{R_{{0}}} \:\mathsf{x}}{12}\right)=0.
\end{equation}
The derivative of Eq. (\ref{vf1}) with respect to $\mathsf{x}$ gives:
\begin{equation}\label{vf2}
\frac{d^2V_{{\rm eff}}}{d\mathsf{x}^2}= \frac{c^2}{\mu^2}\frac{\left(-4\mathsf{R_{{0}}}\mathsf{x}^4+15\mathsf{R_{{0}}}\mathsf{x}^3
+12\mathsf{x}-72\right)}{\mathsf{x}^4\left(-\frac{3}{\mathsf{x}}+1\right)},
\end{equation}
where we have replaced $\mathsf{h}$ by (Harko et al. 2009):
\begin{equation}\label{specificl}
\mathsf{h} = \mathsf{x_{c}}^2 \frac{\sqrt{\frac{1}{\mathsf{x_{c}}^3}-\frac{\mathsf{R_{{0}}}}{12}}}{\sqrt{1-\frac{3}{\mathsf{x_{c}}}}},
\end{equation}
where $\mathsf{x_c}$ corresponds to the radius of a circular orbit. 
%
We have performed the numerical calculation of the extrema of the effective potential for different values of $\mathsf R_0$. 

\begin{figure}[h!]
\center
\resizebox{\hsize}{!}{\includegraphics{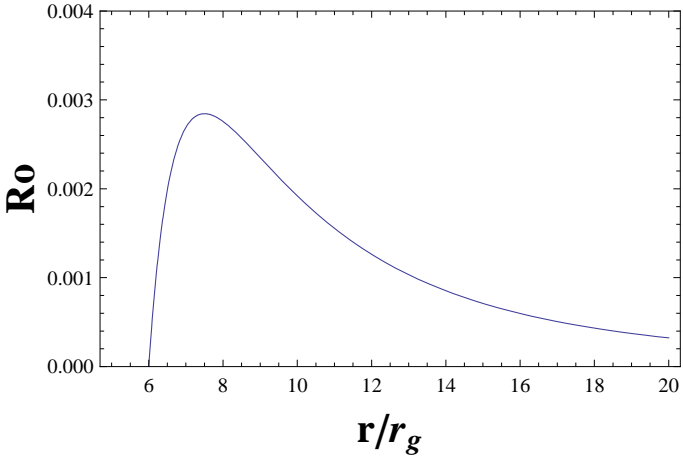}}
\caption{\label{limi-Ro}Plot of the function given by Eq. (\ref{ro-x}). The absolute maximum corresponds to $\mathsf{x} = 15/2$ and $\mathsf{R_{0}} = 2.85\times10^{-3}$.}
\end{figure}

\begin{figure}[h!]
\center
\resizebox{\hsize}{!}{\includegraphics{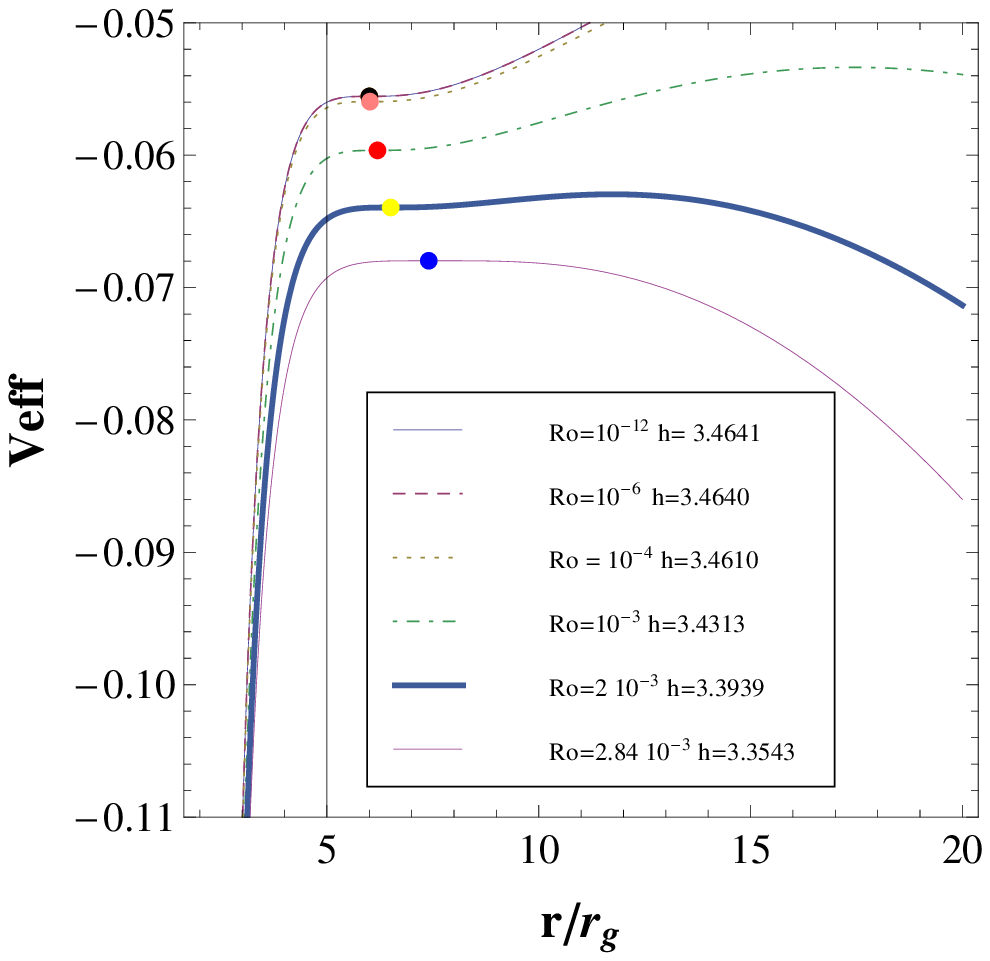}}
\caption{\label{vfpos}Effective potential for different values of $\mathsf{R_0}>0$ and $\mathsf{h}$ in $f(R)$-Schwarzschild space-time. The dots indicate the location of the innermost stable circular orbit.}
\end{figure}

For $\mathsf R_0>0$, as shown by Stuchl{\'\i}k et al. (1999) and Rezzolla et al. (2003) in Schwarzschild-de Sitter space-time, stable circular orbits exist for values of the specific angular momentum that satisfy:
\begin{equation}
h_{\rm isco} < h < h_{\rm osco},
\end{equation}
where $h_{\rm isco}$ and $h_{\rm osco}$ stand for the local minimum and local maximum of the specific angular momentum at the inner and outer marginally stable radii. From Eq. (\ref{vf2}) we see that the existence and location of the circular orbits depend on the Ricci scalar $\mathsf{R_0}$. By equating Eq. (\ref{vf1}) to zero and isolating $\mathsf{R_0}$, we obtain the Ricci scalar as a function of the radial coordinate of the circular orbits:
\begin{equation}\label{ro-x}
\mathsf{R_{0}} = \frac{12 \left(6-\mathsf{x_{c}}\right)}{\left(15-4\mathsf{x_{c}}\right)\mathsf{x_{c}}^{3}}.
\end{equation}
In Figure \ref{limi-Ro} we show the plot of the latter equation. We see that there is an upper limit for the Ricci scalar, $\mathsf{R_{0}} = 2.85\times10^{-3}$, for which circular orbits are possible. In Figure \ref{vfpos} we plot the effective potential as a function of the radial coordinate. The dots indicate the location of the innermost stable circular orbits. The corresponding values of the event and cosmological horizons, radii of the innermost and outermost stable circular orbits, for six different values of $\mathsf{R_{0}} \in (0,2.85\times10^{-3})$ are shown in Table \ref{tab0}. We see that for increasing values of the Ricci scalar the event horizon becomes larger than in Schwarzschild space-time in GR as well as the location of the innermost stable circular orbit.

\begin{table*}
\caption{Location of the event and cosmological horizon, and of the innermost and outermost stable circular orbits for $\mathsf{R_0}>0$ in $f(R)$-Schwarzschild space-time. Here $\mathsf{x}\equiv r/r_{\rm g}$.}
\label{tab0}
\centering
\begin{tabular}{c c c c c}
\hline \hline
\rm $\mathsf{R_0}$ & {\rm Radius}& {\rm Radius} & {\rm Radius innermost} & {\rm Radius outermost}\\ 
 & {event horizon}  & {\rm cosmological horizon} & {\rm stable circular orbit} & {\rm stable circular orbit}\\ \hline
$10^{-12}$ & $\mathsf{x}_{\rm eh}=2$ & $\mathsf{x}_{\rm ch}=3.46\times10^{6}$ & $\mathsf{x}_{\rm isco}=6$ & $\mathsf{x}_{\rm osco}=14421.70$ \\ 
$10^{-6}$ & $\mathsf{x}_{\rm eh}=2$ & $\mathsf{x}_{\rm ch}=3463.10$ & $\mathsf{x}_{\rm isco}=6.00016$  & $\mathsf{x}_{\rm osco}=143.45$\\ 
$10^{-4}$ & $\mathsf{x}_{\rm eh}=2$ & $\mathsf{x}_{\rm ch}=345.40$ & $\mathsf{x}_{\rm isco}=6.02$ & $\mathsf{x}_{\rm osco}=30.16$ \\ 
$10^{-3}$ & $\mathsf{x}_{\rm eh}=2.00067$ & $\mathsf{x}_{\rm ch}=108.53$ & $\mathsf{x}_{\rm isco}=6.19$ & $\mathsf{x}_{\rm osco}=13.17$\\ 
$2\times10^{-3}$ & $\mathsf{x}_{\rm eh}=2.00134$ & $\mathsf{x}_{\rm ch}=76.44$ & $\mathsf{x}_{\rm isco}=6.51$ & $\mathsf{x}_{\rm osco}=9.80$\\
$2.84\times10^{-3}$ & $\mathsf{x}_{\rm eh}=2.0019$ & $\mathsf{x}_{\rm ch}=63.98$ & $\mathsf{x}_{\rm isco}=7.40$ & $\mathsf{x}_{\rm osco}=7.60$\\
\hline
\end{tabular}
\end{table*}

\begin{figure}[h!]
\center
\resizebox{\hsize}{!}{\includegraphics{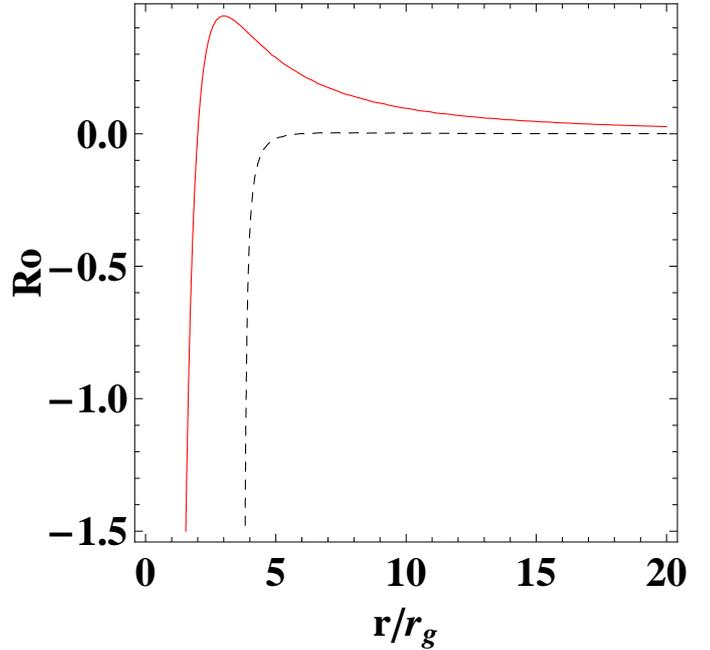}}
\caption{\label{last-orb-Sch1}Plot of Ricci scalar as a function of the radial coordinate of the event horizon (line) and of the Ricci scalar as a function of the radial coordinate of the innermost stable circular orbits (dashed line) for $\mathsf{R_0} \in [-1.5,0.45]$ in $f(R)$-Schwarzschild space-time.}
\end{figure}

\begin{figure}[h!]
\center
\resizebox{\hsize}{!}{\includegraphics{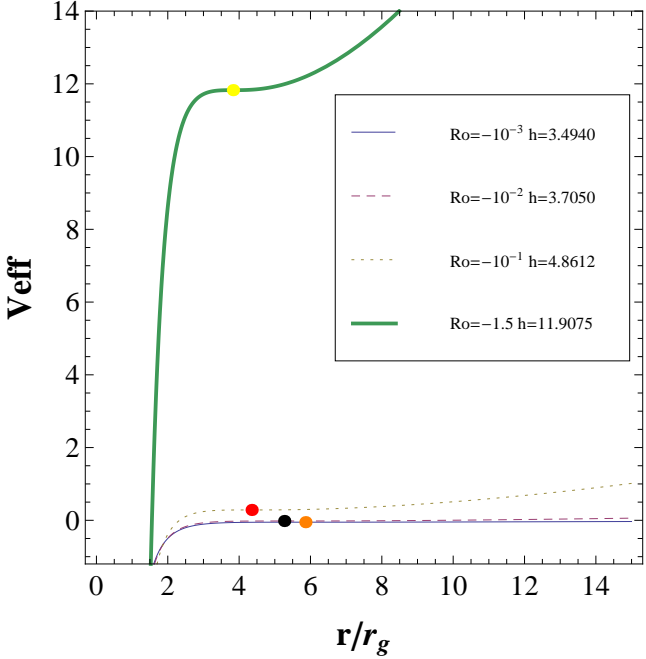}}
\caption{\label{theplot1} Effective potential for different values of $\mathsf{R_0}<0$ and $\mathsf{h}$ in $f(R)$-Schwarzschild space-time. The dots indicate the location of the innermost stable circular orbit.}
\end{figure}

The extrema of the effective potential for $\mathsf{R_{{0}}}<0$ are all located outside the event horizon, as shown in Figure \ref{last-orb-Sch1}. The value of the radial coordinate for the event horizon is less than 2 (i.e. smaller than for Schwarzschild black holes in GR). The location of the innermost stable circular orbit is closer to the horizon than that of the Schwarzschild solution in Einstein's gravity. 
The limit $\mathsf{R_{{0}}} \rightarrow -\infty$ in Eq. (\ref{vf2}) yields: 
\begin{equation}
\mathsf{x}^3\left(-4\mathsf{x}+15\right) = 0 \:\:\:\Rightarrow\:\:\: \mathsf{x}=0 \:\:\vee\:\: \mathsf{x}=3.75.
\end{equation}
The plot of the effective potential corresponding to the values of the parameters of Table \ref{tab-2} is shown in Figure \ref{theplot1}.  
\begin{table}
\caption{Location of the event horizon, and of the innermost stable circular orbit for $\mathsf{R_0}<0$ in $f(R)$-Schwarzschild space-time. Here $\mathsf{x}\equiv r/r_{\rm g}$.}
\label{tab-2}
\centering
\begin{tabular}{c c c}
\hline \hline
\rm $\mathsf{R_0}$ & {\rm Radius}& {\rm Radius innermost} \\ 
 & {event horizon}  & {\rm stable circular orbit}\\ \hline
$-10^{-3}$ & $\mathsf{x}_{\rm eh}=1.999$ & $\mathsf{x}_{\rm isco}=5.86$  \\ 
$-10^{-2}$ & $\mathsf{x}_{\rm eh}=1.993$ & $\mathsf{x}_{\rm isco}=5.26$  \\ 
$-10^{-1}$ & $\mathsf{x}_{\rm eh}=1.939$ & $\mathsf{x}_{\rm isco}=4.35$  \\ 
$-1.5$ & $\mathsf{x}_{\rm eh}=1.541$ & $\mathsf{x}_{\rm isco}=3.83$ \\ 
\hline
\end{tabular}
\end{table}


\subsection{Kerr space-time in $f(R)$ theories}\label{hori-kerr}

The axisymmetric, stationary and constant Ricci scalar
geometry that describes a black hole with mass, electric
charge and angular momentum was found by Carter
(1973), and was used to 
study $f(R)$ black holes by Cembranos et al. (2011). The form of the metric is the following:
\begin{eqnarray}\label{kerr-ftot}
ds^{2}&  = &\frac{\rho^2}{\Delta_{\rm r}} dr^{2}+\frac{\rho^2}{\Delta_{\rm \theta}} {d\theta}^{2}\\ \nonumber
& + & \frac{\Delta_{\rm \theta}\: {\sin^{2}{\theta}}}{\rho^{2}}\left[a\:\frac{c\:dt}{\Xi}-\left(r^{2}+a^{2}\right)\frac{d\phi}{\Xi}\right]^{2}\\ \nonumber
&-&\frac{\Delta_{\rm r}}{\rho^{2}}\left(\frac{c\:dt}{\Xi}-a\:{\sin^{2}{\theta}}\frac{d\phi}{\Xi}\right)^{2},
\end{eqnarray}
where:
\begin{eqnarray}
\Delta_{\rm r}& = &\left(r^{2}+a^{2}\right)\left(1-\frac{R_{{0}}}{12}r^{2}\right)-\frac{2GMr}{c^2},\\
\rho^{2} & = & r^{2}+a^{2}{\cos{\theta}}^{2},\\
\Delta_{\rm \theta}& = & 1+\frac{R_{0}}{12}a^{2}{\cos^{2}{\theta}},\\
\Xi & = & 1+\frac{R_{0}}{12}a^{2}.
\end{eqnarray}
Here $M$ and $a$ denote the mass and angular momentum per unit mass of the black hole, respectively, and $R_0$ is given by Eq.(\ref{condi}).

Because of the constancy of $p^{t}$ and $p^{\phi}$ along the trajectories, 
and of the reflection-symmetry of the metric through the equatorial plane, the orbit of any particle with initial condition $p^{\theta} = 0$ will remain in the plane $\pi/2$, where the metric has the form:
\begin{eqnarray}\label{Kr-metric}
ds^{2} & = & -\frac{c^2}{r^2 \Xi^2} \left(\Delta_{r}-a^{2}\right)dt^{2}+\frac{r^{2}}{\Delta_{r}} dr^{2}\\ \nonumber 
&-&\frac{2ac}{r^{2}\Xi^{2}} \left(r^{2}+a^{2}-\Delta_{r}\right) dt d\phi \\ \nonumber
&+&  \frac{d\phi^{2}}{r^{2}\Xi^{2}} \left[\left(r^{2}+a^{2}\right)^{2}-\Delta_{r} a^{2}\right].
\end{eqnarray}
Here,
\begin{eqnarray}
\Delta_{r}& = &\left(r^{2}+a^{2}\right)\left(1-\frac{R_{{0}}}{12}r^{2}\right)-\frac{2GMr}{c^2},\\
\Xi & = & 1+\frac{R_{0}}{12}a^{2}.
\end{eqnarray}
If $R_{{0}} \rightarrow 0$, Eq. (\ref{Kr-metric}) represents the Kerr space-time metric in GR as expected.

\begin{figure}
\center
\includegraphics[width=8cm]{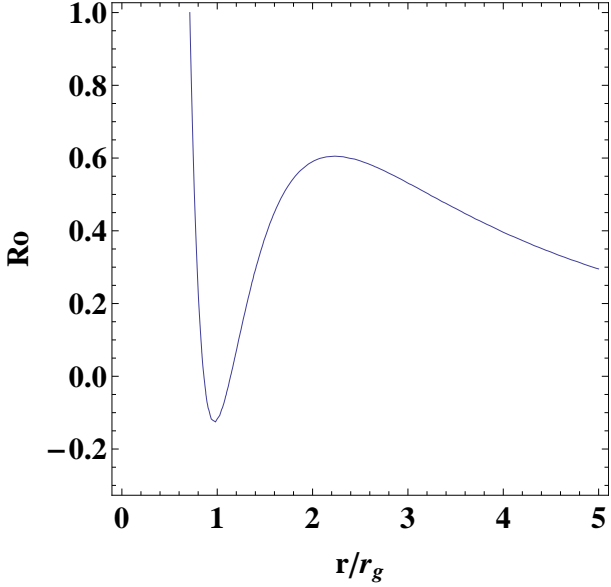}
\caption{\label{or1} Plot of  the Ricci scalar as a function of the radial coordinate of the event horizon for $\mathsf{R_0} \in [-0.3,1]$ and $\mathsf{a} = 0.99$ in $f(R)$-Kerr space-time.}
\end{figure}

The equation that yields the position of the event horizon is obtained by setting $1/g_{{\rm rr}} = 0$:
\begin{equation}
\Delta_{r} =\left(r^{2}+a^{2}\right)\left(1-\frac{R_{{0}}}{12}r^{2}\right)-\frac{2GMr}{c^2} =0.
\end{equation}
In terms of $\mathsf{x}$, $\mathsf{R_0}$, and $\mathsf{a} \equiv a(r_{\rm g})^{-1}
$, this equation takes the form:
\begin{equation}
\left(\mathsf{x}^2+\mathsf{a}^{2}\right)\left(1-\frac{\mathsf{R_0}\mathsf{x}^2}{12}\right)-2\mathsf{x} =0.
\end{equation}
In Figure \ref{or1} we plotted the Ricci scalar as a function of the radial coordinate of the event horizon for $\mathsf{R_0} \in [-0.3,1]$ and  $\mathsf{a} = 0.99$ (i.e. a nearly maximally rotating black hole, such as Cygnus X1, Gou et al. 2011). If $\mathsf{R_0} \in (0,0.6]$, there are 3 event horizons: the inner and outer horizons of the black hole and a cosmological horizon;  for $\mathsf{R_0}>0.6$ there is  a cosmological horizon that becomes smaller for larger values of $\mathsf{R_0}$. If $\mathsf{R_0} \in (-0.13,0)$ there are 2 event horizons. For $\mathsf{R_0} \leq -0.13$ naked singularities occur. In the following we shall analyse the existence of stable circular orbits for $\mathsf{R_ 0} \in (-0.13,0.6]$.

\subsubsection{Equations of motion and effective potential in $f(R)$-Kerr space-time}  
%
%
In order to
obtain an expression for the effective potential, we make
use of the invariant length of the 4-momentum \textbf{p}:
\begin{equation}\label{invar}
g^{\mu \nu} p_{\mu} p_{\nu} = \epsilon^{2},
\end{equation}
where $\epsilon^{2} = c^{2}$ for massive particles and $\epsilon^{2} = 0$ for photons. Since we are only interested in trajectories on the equatorial plane, we set $p_{\theta} =0$, and Eq. (\ref{invar}) gives:
\begin{equation}\label{invar1}
g^{tt} (p_{t})^{2} +2 g^{t \phi} p_{t} p_{\phi} + g^{\phi \phi} {p_{\phi}}^{2} + g^{rr} {p_{r}}^{2} = \epsilon^{2},
\end{equation}
where:
\begin{eqnarray}
p_{t} & = & k c^{2},\label{pt} \\ 
p_{\phi} & = & -h, \label{pphi} \\ 
p_{r} & = & g_{rr} \dot{r}. \label{pr}
\end{eqnarray}
Substituing Eqs. (\ref{pt}), (\ref{pphi}), and (\ref{pr}) into (\ref{invar1}), the equation for ${\dot{r}}^{2}$ takes the form:
\begin{equation}
{\dot{r}}^{2} = g^{rr} \left[\epsilon^{2}-g^{tt} \left(kc^{2}\right)^{2}+ 2 g^{t\phi} hkc^{2}-g^{\phi \phi} h^{2}\right].
\end{equation}
The contravariant components of the space-time metric given by Eq. (\ref{Kr-metric}) are:
\begin{eqnarray}
g^{tt} & = &  \frac{\Xi^{2}}{\Delta_{r} \;c^2 r^{2}}\left[\left(r^{2}+a^{2}\right)^{2}-\Delta_{r}a^{2}\right], \label{gc1} \\ 
g^{rr} & = & -\frac{\Delta_{r}}{r^{2}}, \label{gc2} \\ 
g^{t \phi} & = & \frac{\Xi^{2}}{c \Delta_{r} \;r^{2}} a \left(r^{2}+a^{2}-\Delta_{r}\right), \label{gc3} \\ 
g^{\phi \phi} & = & -\frac{\Xi^{2}}{\Delta_{r}\; r^{2}} \left(\Delta_{r}-a^{2}\right).\label{gc4}
\end{eqnarray}
Hence, the energy equation for a massive particle 
is given by:
\begin{equation}
\frac{1}{2}{\dot{r}}^{2} + V_{\rm eff}(r,a,R_{{0}},k,h)=\frac{c^2}{2}\left(k^{2}-1\right),
\end{equation}
where the effective potential is:
\begin{equation}
V_{\rm eff} (r,a,R_{0},k,h)= \frac{c^2\Delta_{r}}{2 r^{2}} +\frac{c^2}{2}\left(k^{2}-1\right)-\frac{\Xi^{2}}{2r^{4}}\Gamma,
\end{equation}
and:
\begin{equation}
\Gamma \equiv \left[\left(r^{2}+a^{2}\right)ck-a h\right]^{2}-\Delta_{r}\left(ack-h\right)^{2}.
\end{equation}

\subsubsection{Equatorial circular orbits of massive particles}
\label{equat}

If a massive particle is moving in a circular orbit of radius $r_{\rm isco}$, the value of the effective potential at any point of the orbit satisfies the equation:
\begin{equation}\label{cond1}
V_{\rm eff}(r_{\rm isco},a,R_{{0}},k,h) = \frac{c^2}{2}\left(k^{2}-1\right),
\end{equation}
so,
\begin{equation}\label{cond2}
\left. \frac{dV_{\rm eff}}{dr} (r,a,R_{{0}},k,h)\right|_{r_{\rm isco}}=0.
\end{equation}
For a stable orbit, the equation: 
\begin{equation}
\left. \frac{d^{2}V_{\rm eff}}{dr^{2}} (r,a,R_{{0}},k,h)\right|_{r_{\rm isco}} > 0,
\end{equation}
must also be satisfied. As shown in Section \ref{hori-kerr}, there are black holes if $\mathsf{R_0} \in (-0.13,0.6]$.

\begin{figure}
\center
\resizebox{\hsize}{!}{\includegraphics{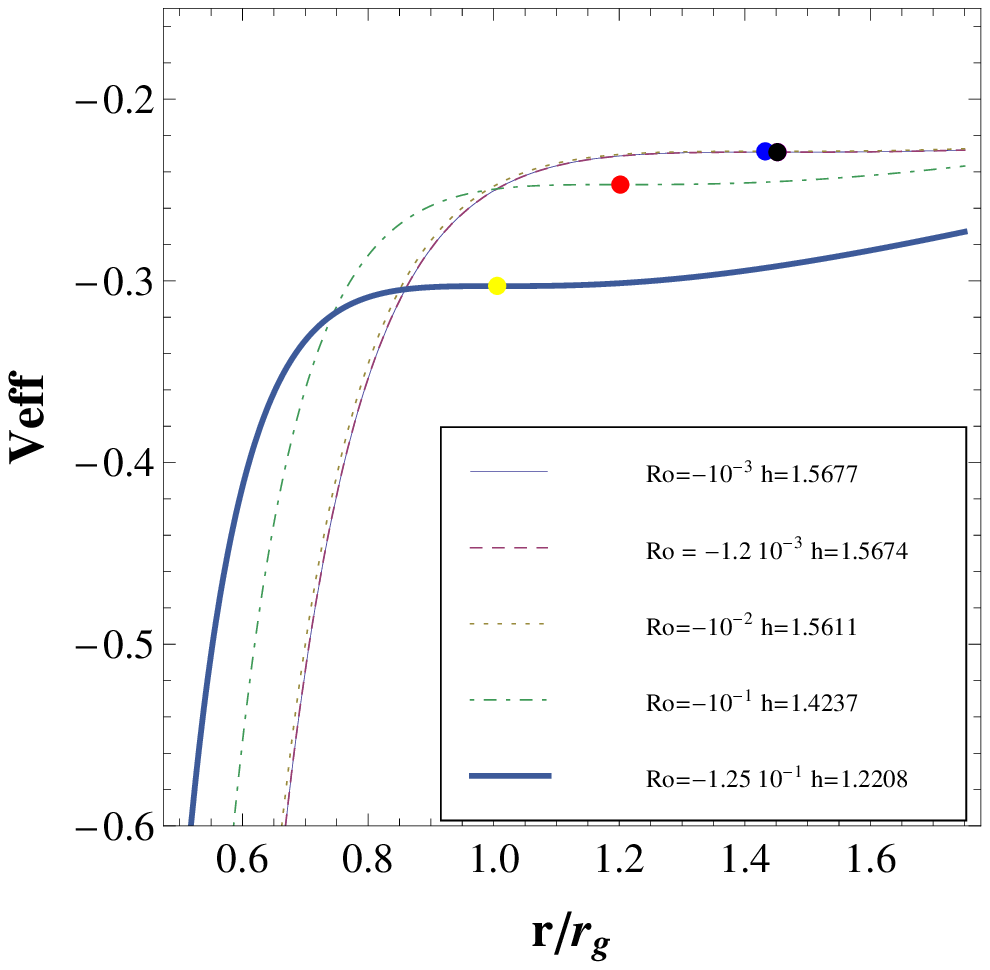}}
\caption{\label{R1}Effective potential as a function of the radial coordinate ($\mathsf{R_{{0}}}<0$, $\mathsf{a} = 0.99$), in $f(R)$-Kerr space-time. The dots indicate the location of the innermost stable circular orbit. }
\label{pot-eff-kerr-1}
\end{figure}

Numerical calculations of the radius of the innermost stable circular orbit for several values of $\mathsf{R_{0}}<0$, with $\mathsf{a}=0.99$, show that the stable circular orbits lay outside the event horizon (see Table \ref{tabla2}). Notice that as the value of the scalar decreases, the radius of the innermost stable circular orbit becomes smaller. The radius of the innermost stable circular orbit in Kerr space-time in GR ($r_{\rm isco} = 1.4545\:r_{\rm g}$, in the prograde case) is always larger than in $f(R)$-Kerr. In Figure \ref{pot-eff-kerr-1}, the effective potential that correspond to the values of Table \ref{tabla2} are shown. 

\begin{figure}
\center
\resizebox{\hsize}{!}{\includegraphics{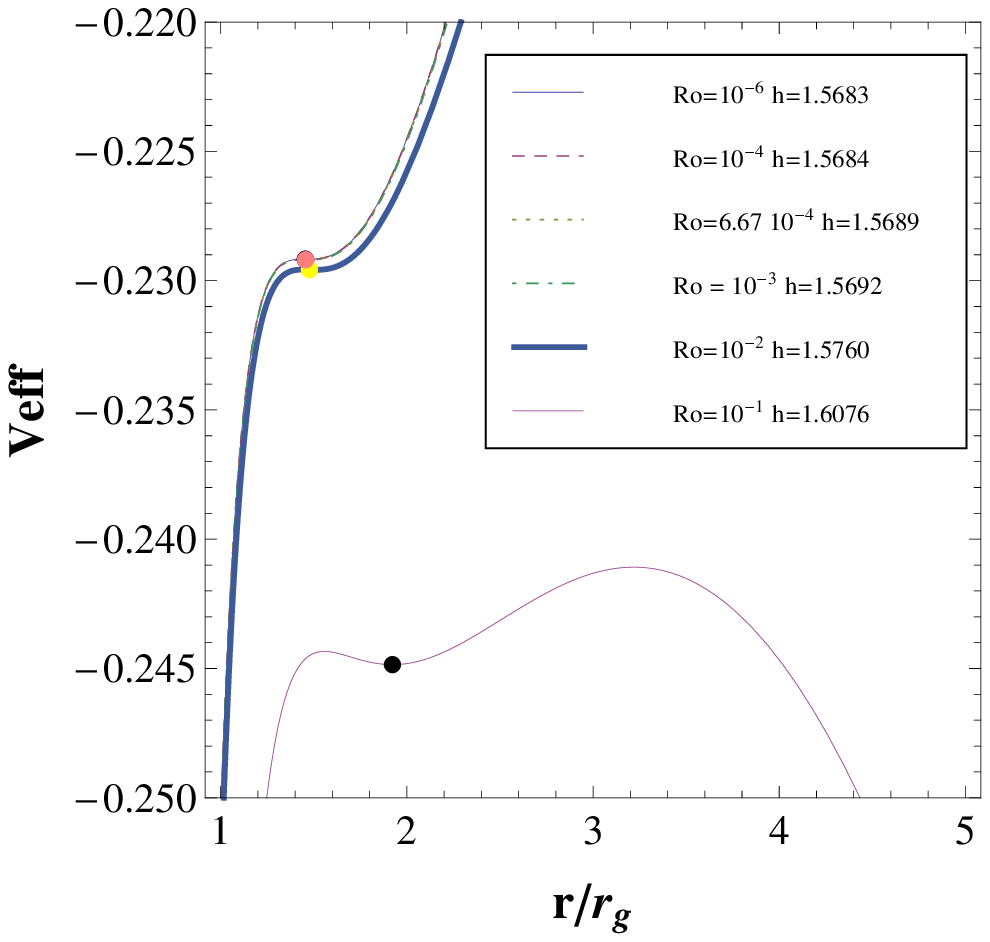}}
\caption{\label{pot-kerr-pos}Effective potential as a function of the radial coordinate ($\mathsf{R_{0}}>0$, $\mathsf{a} = 0.99$), in $f(R)$-Kerr space-time. The dots indicate the location of the innermost stable circular orbit. }
\label{pot-eff-kerr}
\end{figure}

We follow Stuchl{\'\i}k and Slan\'y (2004) to study the existence of stable circular orbits for $\mathsf{R_{0}} \in (0,0.6]$. The specific angular momentum of a massive particle in a co-rotating circular orbit yields (Stuchl{\'\i}k and Slan\'y, 2004):
\begin{equation}\label{hpos}
\mathsf{h} = -\frac{2\mathsf{a}+\mathsf{a}\mathsf{x_c}\left(\mathsf{x_c}^{2}+\mathsf{a}^{2}\right)\frac{\mathsf{R_{0}}}{12}-\mathsf{x_c}\left(\mathsf{x_c}^{2}+\mathsf{a}^{2}\right)\left(\frac{1}{\mathsf{x_c}^{3}}-\frac{\mathsf{R_{0}}}{12}\right)^{1/2}}{\mathsf{x_c}\left[1-\frac{3}{\mathsf{x_c}}-\frac{\mathsf{a}^{2}R_{0}}{12}+2\mathsf{a}\left(\frac{1}{\mathsf{x_c}^{3}}-\frac{\mathsf{R_0}}{12}\right)^{1/2}\right]^{1/2}}.
\end{equation} 
We see from the latter equation that circular orbits must satisfy the following two conditions:
\begin{equation}
\mathsf{x_c} < \left(\frac{12}{\mathsf{R_{0}}}\right)^{1/3},
\end{equation}
which is the same for $f(R)$-Schwarzschild space-time with positive Ricci scalar, and:
\begin{equation}
1-\frac{3}{\mathsf{x_c}}-\frac{\mathsf{a}^{2}\mathsf{R_{0}}}{12}+2\mathsf{a}\left(\frac{1}{\mathsf{x_c}^{3}}-\frac{\mathsf{R_0}}{12}\right)^{1/2}\geq 0.
\end{equation}
The minimum and maximum of Eq. (\ref{hpos}) give the values of the specific angular momentum that correspond to the innermost and outermost stable circular orbit respectively, once the angular momentum $\mathsf{a}$ of the black hole and $\mathsf{R_0}$ are fixed. We show in Table \ref{tab4} the values of such radii for different values of $\mathsf{R_0}$, and $\mathsf{a} = 0.99$, and in Figure \ref{pot-kerr-pos} the corresponding plot of the effective potential. As expected, the radius of inner circular orbit is larger than in Kerr space-time in GR. We also found, by equating to zero the derivative of Eq. (\ref{hpos}), that stable circular orbits only exist if $\mathsf{R_{0}} \in (0,1.45\times10^{-1})$.

\begin{table}
\caption{Radii of event horizons and circular orbits for a $f(R)$-Kerr black hole of angular momentum $\mathsf{a}=0.99$, for some values of $\mathsf{R_0}<0$. Here $\mathsf{x}\equiv r/r_{{\rm g}}$.}
\label{tabla2}
\centering
\begin{tabular}{c c c}
\hline \hline
$\mathsf{R_0}$& {\rm Radii event horizons} & {\rm Radius innermost } \\ 
 &  & stable circular orbits\\ \hline
$-10^{-3}$ & $\mathsf{x}_{\rm eh1}=0.86$, $\mathsf{x}_{\rm eh2}=1.14$ & $\mathsf{x}_{\rm isco} = 1.452$\\ 
$-1.2\times10^{-3}$ & $\mathsf{x}_{\rm eh1}=0.86$, $\mathsf{x}_{\rm eh2}=1.14$ & $\mathsf{x}_{\rm isco} = 1.451$\\ 
$-10^{-2}$ & $\mathsf{x}_{\rm eh1}=0.86$, $\mathsf{x}_{\rm eh2}=1.13$ & $\mathsf{x}_{\rm isco} = 1.43$\\ 
$-10^{-1}$ & $\mathsf{x}_{\rm eh1}=0.91$, $\mathsf{x}_{\rm eh2}=1.03$ & $\mathsf{x}_{\rm isco}=1.20$\\ 
$-1.25\times10^{-1}$ & $\mathsf{x}_{\rm eh1}=0.96$, $\mathsf{x}_{\rm eh2}=0.98$ & $\mathsf{x}_{\rm isco}= 1.04$\\ 
\hline
\end{tabular}
\end{table}

\begin{table*}
\caption{Location of the event and cosmological horizon, and of the innermost and outermost stable circular orbits for $\mathsf{R_0}>0$ and $\mathsf{a}=0.99$ in $f(R)$-Kerr space-time. Here $\mathsf{x}\equiv r/r_{\rm g}$.}
\label{tab4}
\centering
\begin{tabular}{c c c c c}
\hline \hline
\rm $\mathsf{R_0}$ & {\rm Radii}& {\rm Radius} & {\rm Radius innermost} & {\rm Radius outermost}\\ 
 & {event horizons}  & {\rm cosmological horizon} & {\rm stable circular orbit} & {\rm stable circular orbit}\\ \hline
$10^{-6}$ & $\mathsf{x}_{\rm eh1}=0.86$ $\mathsf{x}_{\rm eh2}=1.14$& $\mathsf{x}_{\rm ch}=3463$ & $\mathsf{x}_{\rm isco}=1.4545$  & $\mathsf{x}_{\rm osco}=143.59$\\ 
$10^{-4}$ & $\mathsf{x}_{\rm eh1}=0.86$ $\mathsf{x}_{\rm eh2}=1.14$& $\mathsf{x}_{\rm ch}=345.40$ & $\mathsf{x}_{\rm isco}=1.4547$ & $\mathsf{x}_{\rm osco}=30.53$\\
$6.67\times10^{-4}$ & $\mathsf{x}_{\rm eh1}=0.86$ $\mathsf{x}_{\rm eh2}=1.14$& $\mathsf{x}_{\rm ch}=113.12$ & $\mathsf{x}_{\rm isco}=1.4560$ & $\mathsf{x}_{\rm osco}=16.002$\\ 
$10^{-3}$ & $\mathsf{x}_{\rm eh1}=0.86$ $\mathsf{x}_{\rm eh2}=1.14$& $\mathsf{x}_{\rm ch}=108.53$ & $\mathsf{x}_{\rm isco}=1.4567$ & $\mathsf{x}_{\rm osco}=13.93$\\ 
$10^{-2}$ & $\mathsf{x}_{\rm eh1}=0.85$ $\mathsf{x}_{\rm eh2}=1.15$& $\mathsf{x}_{\rm ch}=33.59$ & $\mathsf{x}_{\rm isco}=1.4765$ & $\mathsf{x}_{\rm osco}=6.25$\\
$10^{-1}$ & $\mathsf{x}_{\rm eh1}=0.83$ $\mathsf{x}_{\rm eh2}=1.22$& $\mathsf{x}_{\rm ch}=9.80$ & $\mathsf{x}_{\rm isco}=1.92$ & $\mathsf{x}_{\rm osco}=3.22$\\
\hline
\end{tabular}
\end{table*}

%

The analysis of the circular orbits presented in this section 
will be applied next to the construction of accretion disks around
black holes.  


\section{Accretion disks in strong gravity}
\label{sacc}

\subsection{Standard disk model in general relativity}

The first realistic model of accretion disks around black holes was formulated by Shakura and Sunyaev (1973). They considered that the matter rotating in circular Keplerian orbits around the compact object loses angular momentum  
because of the friction between adjacent layers and spirals inwards. In the process gravitational energy is released, the kinetic energy of the plasma increases and the disk heats up, emitting thermal energy.

Novikov, Thorne, and Page (Novikov \& Thorne 1973; Page \& Thorne 1974) made a relativistic analysis of the structure of an accretion disk around a black hole. They assumed the background space-time geometry to be stationary, axially-symmetric, asymptotically flat, and reflection-symmetric with respect to the equatorial plane. They also postulated that the central plane of the disk coincides with the equatorial plane of the black hole. This assumption entails that the metric coefficients  $g_{ tt}$, $g_{t\phi}$, $g_{rr}$, $g_{\theta \theta}$, and $g_{ \phi \phi}$ depend only on the radial coordinate $r$.

The disk is supposed to be in a quasi-steady state (Novikov \& Thorne 1973), so any relevant quantity (for example the density or the temperature of the gas) is averaged over $2\pi$, a proper radial distance of order $2H$\footnote{Here $H$ represents a particular height above the central plane of the disk ($\left|z\right|\leq H << r$).}, and the time interval $\Delta t$ that the gas takes to move inward a distance $2H$. During $\Delta t$, the changes in the space-time geometry are negligible.

Particles move in the equatorial plane in nearly geodesic orbits; consequently, the gravitational forces exerted by the black hole completely dominate over the radial accelerations due to pressure gradients.

The expression of the energy flux for a relativistic accretion disk takes the form (Novikov \& Thorne 1973; Page \& Thorne 1974):
\begin{equation}\label{flux}
Q(r) = - \frac{\dot{M_{0}}}{4\pi \sqrt{-g}}\frac{\Omega,_{r}}{\left(\widetilde{E}-\Omega\widetilde{L}\right)^{2}}
\int^{r}_{r_{{\rm isco}}} \left(\widetilde{E}-\Omega\widetilde{L}\right) \widetilde{L,_{r}} dr,
\end{equation}
where $\dot{M_{0}}$ stands for the mass accretion rate, $\Omega$ for the angular velocity and $\widetilde{E}$ and $\widetilde{L}$ represent the specific energy and angular momentum, respectively. The lower limit of the integral $r_{{\rm isco}}$ corresponds to the location of the innermost stable circular orbit. 

The angular velocity $\Omega$, the specific energy $\widetilde{E}$, and the specific angular momentum $\widetilde{L}$ of the particles moving in circular orbits are given by (Harko et al. 2009):
\begin{eqnarray}
\Omega &  = & \frac{d\phi}{dt} = \frac{-g_{t\phi,r}+\sqrt{\left(-g_{t\phi,r}\right)^{2}-g_{tt,r}g_{\phi\phi,r}}}{g_{\phi\phi,r}},\label{omega1}\\
\widetilde{E} & = & - \frac{g_{tt} + g_{t\phi} \Omega}{\sqrt{-g_{tt}-2g_{t\phi}\Omega-g_{\phi\phi}\Omega^{2}}},\label{energia}\\
\widetilde{L} & = & \frac{g_{t\phi}+g_{\phi\phi} \Omega}{\sqrt{-g_{tt}-2g_{t\phi}\Omega-g_{\phi\phi}\Omega^{2}}}.\label{moml}
\end{eqnarray}
Eqs. (\ref{energia}), and (\ref{moml}) can be derived by writing the effective potential $V_{\rm eff}(r)$ in terms of the metric coefficients and solving for $\widetilde{E}$ and $\widetilde{L}$ the equations $V_{\rm eff}(r) = 0$ and ${V_{\rm eff}}_{,{\rm r}}(r) = 0$. The formula for the angular velocity $\Omega = d\phi/dt$ is obtained by substituing $\widetilde{E}$, and $\widetilde{L}$ into the geodesic equations $dt/d\tau$ and $d\phi/d\tau$ (Harko et al. 2009).

In the next subsections we calculate the energy flux, temperature and luminosity of an accretion disk around a Schwarzschild and a Kerr black hole in GR and $f(R)$ gravity with constant Ricci scalar, 
adopting the following values for the relevant parameters:
$M=14.8 M_{\odot}$, $\dot{M}= 0.472\times10^{19} \rm{g}\:\rm{s}^{-1}$, and $\mathsf{a} = 0.99$, which are the best estimates available for 
the well-known galactic black hole Cygnus-X1 (Orosz et al. 2011, Gou et al. 2011).
 
\subsubsection{Relativistic accretion disk around Schwarzschild and Kerr black holes}

In order to obtain an expression of the energy flux and temperature of the disk, for the Schwarzschild black hole, we calculate the angular velocity $\Omega$, the specific energy $\widetilde{E}$ and angular momentum $\widetilde{L}$ of the particles in the disk, using the metric (Schwarzschild 1916):
\begin{eqnarray}\label{gsch}
ds^{2} &=& - c^{2} \left(1-\frac{2GM}{c^{2}r}\right) dt^{2} + \left(1-\frac{2GM}{c^{2}r}\right)^{-1} dr^{2}\\ \nonumber 
&+& r^{2} d\theta^{2} + r^{2} \sin\theta^{2} d\phi^{2}.
\end{eqnarray}
From Eqs. (\ref{omega1}), (\ref{energia}), and (\ref{moml}) we get: 
\begin{eqnarray}
\Omega &  = & \sqrt{\frac{GM}{r^{3}}},\label{omega} \\ 
\widetilde{E} & = & c \frac{\left(1-\frac{2GM}{c^{2}r}\right)}{\sqrt{\left(1-\frac{3GM}{c^{2}r}\right)}}, \label{e} \\ 
\widetilde{L} & = & \frac{\sqrt{GM} \sqrt{r}}{c \sqrt{\left(1-\frac{3GM}{c^{2}r}\right)}}.\label{l}
\end{eqnarray} 
By replacing these equations 
in Eq. (\ref{flux}), we obtain:
\begin{eqnarray}\label{fluxsch}
Q(\mathsf{x})& =& \frac{3 \dot{M_{0}} c^{6}}{8 \pi \mathsf{x}^{7/2}}  \frac{1}{\left(GM\right)^{2}}\left(1-\frac{3}{\mathsf{x}}\right)^{-1} \times\\ \nonumber
& &\left[\sqrt{\mathsf{x}}+\sqrt{3}\tanh^{-1}\sqrt{\frac{\mathsf{x}}{3}}\;\right]^{\mathsf x}_{\mathsf{x_{{\rm isco}}}},
\end{eqnarray}
where 
$\mathsf{x_{{\rm isco}}}= 6r_{\rm g}$ is the location of the innermost stable circular orbit in Schwarzschild space-time.

By means of Stefan-Bolzmann's law, 
\begin{equation}
T(r) = z\left(\frac{Q(r)}{\sigma_{\rm SB}}\right)^{1/4},
\end{equation}
(where $z$ stands for the correction due to the gravitational redshift),
the 
temperature of the disk 
can be obtained 
as a function of the radial coordinate $r$.

\begin{figure}[t]
\center
\resizebox{\hsize}{!}{\includegraphics{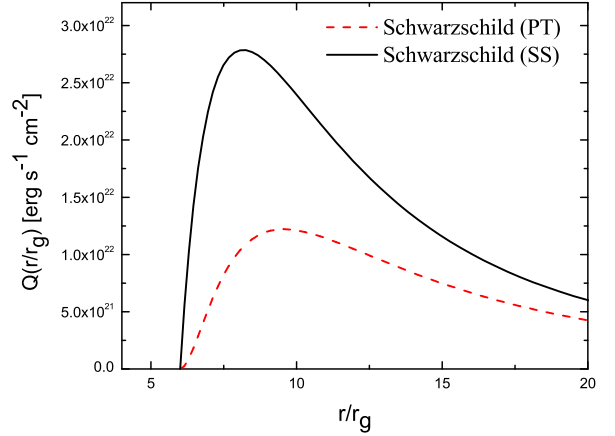}}
\caption{Energy flux as a function of the radial coordinate of an accretion disk around a Schwarzschild black hole in Shakura-Sunyaev (SS) and Page-Thorne (PT) models,  respectively.}
\label{flujo-ss}
\end{figure}

\begin{figure}[t]
\center
\resizebox{\hsize}{!}{\includegraphics{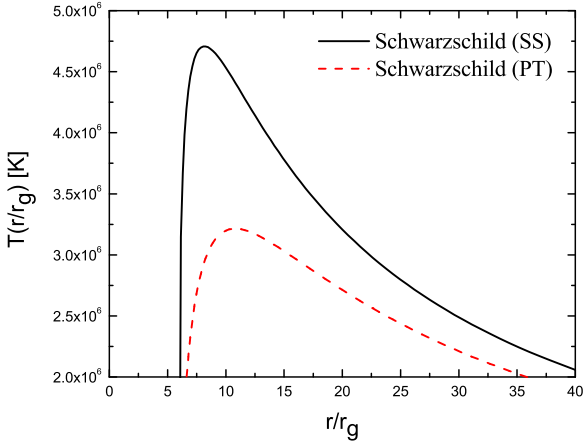}}
\caption{Temperature as a function of the radial coordinate of an accretion disk around a Schwarzschild black hole in SS and PT models, respectively.}
\label{temp-ss}
\end{figure}

In Figures \ref{flujo-ss} and \ref{temp-ss} we plot the energy flux and the temperature as a function of the radial coordinate for a Schwarzschild black hole in both Keplerian and relativistic accretion disk models. We see that GR effects introduce a decrease of the peak of the energy flux by a factor $\approx$ 2, and that the temperature distribution is also diminished.

The Kerr space-time metric (Kerr 1963) in Boyer-Lindquist coordinates for 
$\theta=\pi/2$ is:
\begin{eqnarray}\label{K-metric}
ds^{2} &=& -\frac{c^2}{r^2} \left(\Delta_{r}-a^{2}\right)dt^{2}+\frac{r^{2}}{\Delta_{r}} dr^{2}\\ \nonumber
& -&\frac{2ac}{r^{2}} \left(r^{2}+a^{2}-\Delta_{r}\right) dt d\phi \\ \nonumber
&  
+ & \frac{d\phi^{2}}{r^{2}} \left[\left(r^{2}+a^{2}\right)^{2}-\Delta_{r} a^{2}\right],
\end{eqnarray}
where: 
\begin{equation}
\Delta_{r} \equiv \left(r^{2}+a^{2}\right)-\frac{2GMr}{c^2}.
\end{equation}
The expression of the energy flux now becomes:
\begin{equation}
Q(r) = - \frac{\dot{M_{0}}}{4\pi \sqrt{-g}}\frac{\Omega,_{r}}{\left(\widetilde{E}-\Omega\widetilde{L}\right)^{2}}
\int^{r}_{r_{{\rm isco}}} \left(\widetilde{E}-\Omega\widetilde{L}\right) \widetilde{L,_{r}} dr,
\end{equation}
where:
\begin{equation}
\left(\widetilde{E}-\Omega\widetilde{L}\right) \widetilde{L,_{r}} = - \frac{c}{2}\frac{\left[\left(p_{1}+p_{2}\right)\:p_{3}\right]}{p_{4}},
\end{equation}
\begin{equation}
- \frac{\dot{M_{0}}}{4\pi \sqrt{-g}}\frac{\Omega,_{r}}{\left(\widetilde{E}-\Omega\widetilde{L}\right)^{2}} = \frac{3\dot{M_0}}{8\pi}\frac{c^{2}}{\mu^{2}}\frac{p_{5}}{\left(p_{1}+p_{2}\right)^{2}},
\end{equation}
and the coefficients $p_i$ are given by:
\begin{eqnarray}
p_{1} & = & x^{5/2}\left(1-\frac{2}{\mathsf{x}}+\frac{\mathsf{a}}{\mathsf{x}^{3/2}}\right)
\left(-1+\frac{\mathsf{a}^{2}}{\mathsf{x}^{3}}\right),\\
p_{2} & = &  \left(-\mathsf{a}+\mathsf{x}^{3/2}\right)\left(1+\frac{\mathsf{a}^{2}}{\mathsf{x}^{2}}
-2\frac{\mathsf{a}}{\mathsf{x}^{3/2}}\right),\\
p_{3} & = & \left(1+\frac{\mathsf{a}}{\mathsf{x}^{3/2}}\right)\left(-1+\frac{6}{\mathsf{x}}-
\frac{8\mathsf{a}}{\mathsf{x}^{3/2}}+\frac{3\mathsf{a}^{2}}{\mathsf{x}^{2}}\right),\\
p_{4} & = & \mathsf{x}^{3}\:\left(1-\frac{3}{\mathsf{x}}+\frac{2\mathsf{a}}{\mathsf{x}^{3/2}}\right)^{2}
\left(-1+\frac{\mathsf{a}^{2}}{\mathsf{x}^{3}}\right),\\
p_{5} & = & \frac{1}{\mathsf{x}^{3/2}}\:\left(\mathsf{x}^{3}+\mathsf{a}-2\mathsf{a}\mathsf{x}^{3/2}\right)
\left(1-\frac{3}{\mathsf{x}}+\frac{2\mathsf{a}}{\mathsf{x}^{3/2}}\right).
\end{eqnarray}
Here $\mathsf{x} = r/r_{\rm g}$ is an adimensional radial coordinate, and $\mathsf{a}= a/r_{\rm g}$ is the angular momentum of the black hole in adimensional units. 

\begin{figure}[t]
\center
\resizebox{\hsize}{!}{\includegraphics{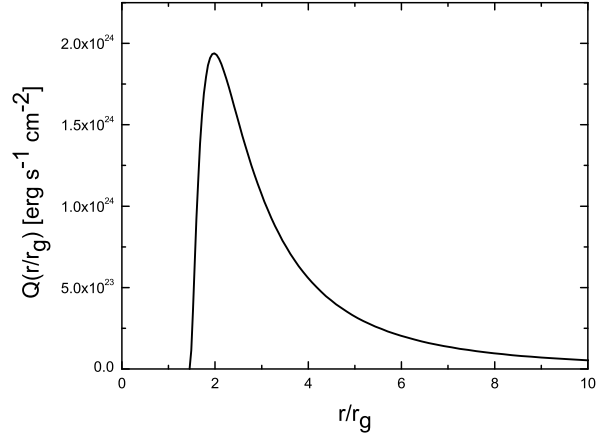}}
\caption{Energy flux as function of the radial coordinate for an accretion disk around a Kerr black hole of angular momentum $\mathsf{a}=0.99$ in the PT model.}
\label{kerr1}
\end{figure}

\begin{figure}[t]
\center
\resizebox{\hsize}{!}{\includegraphics{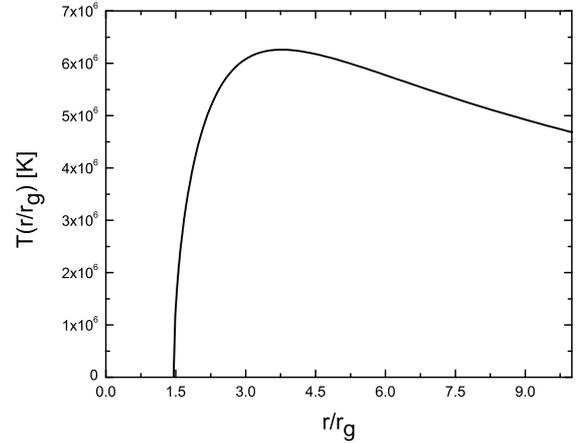}}
\caption{Temperature as function of the radial coordinate of an accretion disk around a Kerr black hole of angular momentum $\mathsf{a}=0.99$ in the PT model.}
\label{kerr2}
\end{figure}

\begin{figure}[t]
\center
\resizebox{\hsize}{!}{\includegraphics{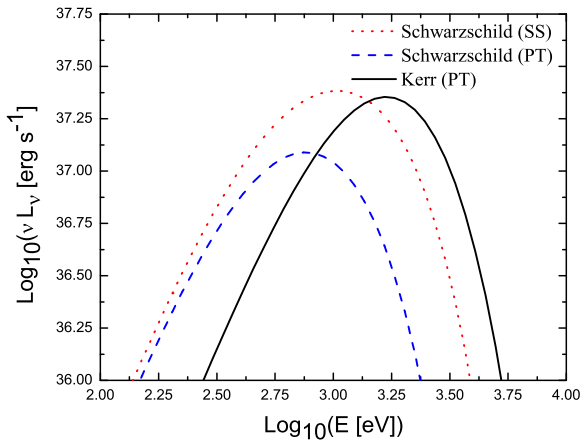}}
\caption{Luminosity as function of the energy for a relativistic accretion disk around a Schwarzschild and a Kerr black hole ($\mathsf{a}=0.99$). We also plot the luminosity as a function of the energy of an accretion disk around a Schwarzschild black hole that corresponds to the SS model.}
\label{kerr3}
\end{figure}

In Figures \ref{kerr1} and \ref{kerr2} we show the plots of the energy flux and temperature of an accretion disk around a Kerr black hole of angular momentum $\mathsf{a}=0.99$, whose innermost stable circular orbit is at $r_{\rm isco} = 1.4545 \:r_{\rm g}$. In Figure \ref{kerr3} we
show the luminosity of relativistic accretion disks around both Schwarzschild and Kerr black holes. For comparison, we also present the Schwarzschild/Shakura-Sunyaev luminosity. 

The values of the maximum temperature, luminosity, and the energy of the peak of the emission for all these models 
are shown in Table \ref{tablavalgr}. As expected, the highest luminosity corresponds to a prograding accretion disk around a 
 black hole. Since the last stable circular orbit is located closer to the black hole than in Schwarzschild space-time, thermal radiation is emitted at higher energies.

\begin{table*}
\caption{Values of the energy of the peak of the emission, the maximum temperature, and luminosity of an accretion disk around Schwarzschild and Kerr black holes ($\mathsf{a}=0.99$) in the SS and PT models.}
\label{tablavalgr}
\centering
\begin{tabular}{c c c c}
\hline\hline
 &  {\rm Schwarzschild (SS)}& {\rm Schwarzschild (PT)} & {\rm Kerr (PT)} \\ 
\hline
$E_{{\rm max}}$ & $1007.59\: {\rm eV}$ & $746.94\: {\rm eV}$  & $1654.9\: {\rm eV}$\\ 
$T_{{\rm max}}$ & $0.406\:{\rm keV}$ & $0.277\:{\rm keV}$ & $0.539\:{\rm keV}$\\ 
$L(E_{{\rm max}})$ & $2.42\times10^{37}\:{\rm erg\:s}^{-1}$ & $1.22\times10^{37}\:{\rm erg\:s}^{-1}$ & $2.26\times10^{37}\:{\rm erg\:s}^{-1}$\\ 
\hline
\end{tabular}
\end{table*}


\subsection{$f(R)$-gravity}

\subsubsection{$f(R)$-Schwarzschild black holes}

The energy flux of an accretion disk around a $f(R)$-Schwarzschild black hole with metric given by Eq. (\ref{g-Sc}) takes the form:
\begin{equation}
Q = \frac{9\dot{M_0}c^{6}}{4\pi\left(GM\right)^{2}}\frac{\left(1-\frac{3}{\mathsf{x}}\right)^{-1}}
{\mathsf{x}^{5}\sqrt{\frac{36}{\mathsf{x}^{3}}-3c^{2}\mathsf{R_0}}}\int^{\mathsf{x}}_{\mathsf{x}_{\rm isco}}\left(\widetilde{E}-\Omega\widetilde{L}\right)\widetilde{L}_{,{\rm r}} d\mathsf{x},
\end{equation}
where
\begin{equation}
\left(\widetilde{E}-\Omega\widetilde{L}\right)\widetilde{L}_{,{\rm r}} =-\frac{\sqrt{3}}{12\mathsf{x}^{3}}\frac{\left(-12\mathsf{x}+72+
4\mathsf{R_0}\mathsf{x}^{4}-15\mathsf{R_0}\mathsf{x}^{3}\right)}{\left(1-\frac{3}{\mathsf{x}}
\right)\left(\frac{12}{\mathsf{x}^{3}}-\mathsf{R_0}\right)^{1/2}}.
\end{equation}
We adopt for the radius of the outer edge of the disk (Dove et al. 1997):
\begin{equation}\label{re}
r_{\rm out} = 11 r_{\rm isco}.
\end{equation}
According to the latter equation, if we take for the innermost stable circular orbit $r_{\rm isco} = 6.3\:r_{\rm g} $, the outer egde of the disk yields approximately $70 \:r_{\rm g}$. If larger values of $r_{\rm out}$ are considered, there are no major differences in the temperature and luminosity distributions.

\begin{figure}[h!]
\center
\resizebox{\hsize}{!}{\includegraphics{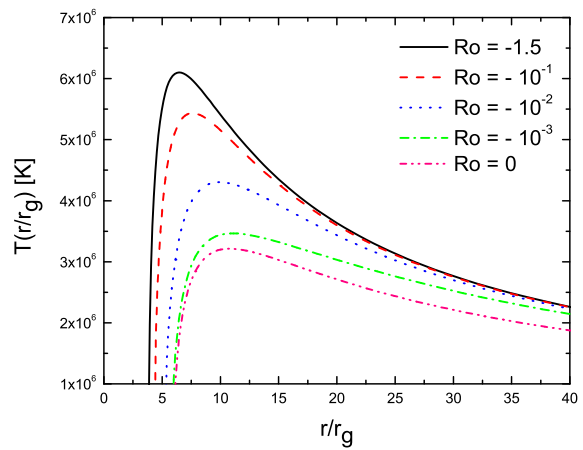}}
\caption{\label{tempsf1}Plot of the temperature as a function of the radial coordinate  for some values of $\mathsf{R_0}< 0 $, for a $f(R)$-Schwarzschild black hole.}
\end{figure}

\begin{figure}[h!]
\center
\resizebox{\hsize}{!}{\includegraphics{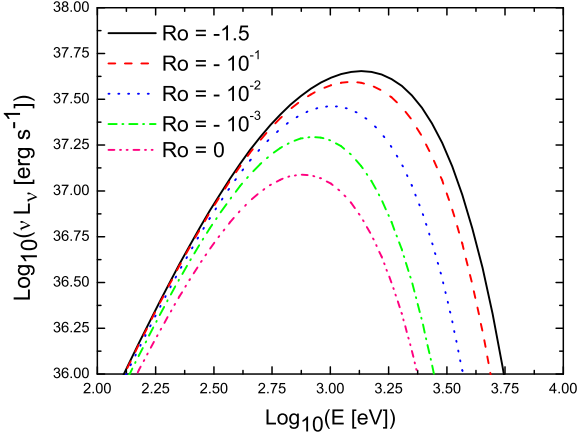}}
\caption{\label{lumisf1}Plot of the luminosity as a function of the energy for some values of $\mathsf{R_0} <0$, for a $f(R)$-Schwarzschild black hole.}
\end{figure}

We first compute the temperature and luminosity spectra distributions for $\mathsf{R_{0}} <0$, and adopt the values given in Table \ref{tab-2}. In Figures \ref{tempsf1} and \ref{lumisf1} we show the plots of the temperature as a function of the radial coordinate, and of the luminosity as a function of the energy, respectively. Notice that the corrections due to the gravitational redshift have been taken into account.

The maximum temperature as well as the luminosity increase for smaller values of $\mathsf{R_{0}}$ (Figures \ref{tempsf1} and \ref{lumisf1}). In the four cases presented, the accretion disk is hotter than in GR, e.g. in a $f(R)$-Schwarzschild disk for $\mathsf{R_{0}} = -1.5$, the maximum temperature and luminosity  are a factor 1.9 and 3.7, respectively, higher. 
The energy corresponding to the peak of the emission shifts to higher values,  reaching 1359.20 eV for the set of adopted parameters. 


\begin{table*}
\caption{Values of the location of the last stable circular orbit, location in the radial coordinate of the maximum temperature, maximum temperature and luminosity, and the energy of the peak of the emission for an accretion disk around a $f(R)$-Schwarzschild black hole with $\mathsf{R_0}<0$.}
\label{tabla3}
\centering
\begin{tabular}{c c c c c c}
\hline\hline
\rm $f(R)$-Schwarzschild  &  $\mathsf{R_0}=0$ & $\mathsf{R_0}=-10^{-3}$ & $\mathsf{R_0}=-10^{-2}$ & $\mathsf{R_0}=-10^{-1}$ & $\mathsf{R_0}=-1.5$ \\
\hline
$r_{{\rm isco}}/r_{{\rm g}}$ & $6$ & $5.85$ & $5.26$ & $4.35$ & $3.82$ \\ 
$r_{{\rm {\tiny Tmax}}}/r_{{\rm g}}$ & $10.82$ & $11.04$  & $10.05$ & $7.58$ & $6.45$\\ 
$T_{{\rm max}}$ & $0.277\:{\rm keV}$ & $0.298\:{\rm keV}$ & $0.371\:{\rm keV}$ & $0.468\:{\rm keV}$ & $0.526\:{\rm keV}$\\ 
$E_{{\rm max}}$ & $746.94\:{\rm eV}$ & $825.31\:{\rm eV}$ & $1007.59\:{\rm eV}$ & $1230.13\:{\rm eV}$ & $1359.20\:{\rm eV}$ \\ 
$L(E_{{\rm max}})$ & $1.22\times10^{37}\:{\rm erg\:s}^{-1}$ & $1.97\times10^{37}\:{\rm erg\:s}^{-1}$ & $2.91\times10^{37}\:{\rm erg\:s}^{-1}$ & $3.93\times10^{37}\:{\rm erg\:s}^{-1}$ & $4.5\times10^{37}\:{\rm erg\:s}^{-1}$\\ 
\hline
\end{tabular}
\end{table*}

We showed in Section \ref{orbits-sch} that for $\mathsf{R_0} >0$, stable circular orbits are possible within a minimum and maximum radius. We see in Table \ref{tab0} that only for $\mathsf{R_0} = 10^{-12}$ and $\mathsf{R_0} = 10^{-6}$ accretion disks are possible, if we take for the radius of the outer edge of the disk $r_{\rm out} = 70r_{\rm g}$. The values of the location of the innermost stable circular orbit, location in the radial coordinate of the maximum temperature, maximum temperature and luminosity, and the energy of the peak of the emission are displayed in Table \ref{tabla3}. We conclude that for $\mathsf{R_{0}} \in (0,10^{-6}]$ the temperature and energy distribution have no significant differences with Schwarzschild's distributions in GR.


\begin{table*}
\caption{Values of the location of the innermost stable circular orbit, location in the radial coordinate of the maximum temperature, maximum temperature and luminosity, and the energy of the peak of the emission for an accretion disk around a $f(R)$-Schwarzschild black hole with $\mathsf{R_0}>0$.}
\label{tabla3}
\centering
\begin{tabular}{c c c}
\hline\hline
\rm $f(R)$-Schwarzschild  &  $\mathsf{R_0}=0$  & $\mathsf{R_0}=10^{-6}$\\
\hline
$r_{{\rm isco}}/r_{{\rm g}}$ & $6$  & $6.00016$ \\ 
$r_{{\rm {\tiny Tmax}}}/r_{{\rm g}}$ & $10.81$   & $10.82$ \\ 
$T_{{\rm max}}$ & $0.277\:{\rm keV}$ & $0.277\:{\rm keV}$ \\ 
$E_{{\rm max}}$ & $746.94\:{\rm eV}$  & $746.94\:{\rm eV}$  \\ 
$L(E_{{\rm max}})$ & $1.22\times10^{37}\:{\rm erg\:s}^{-1}$ & $1.22\times10^{37}\:{\rm erg\:s}^{-1}$\\ 
\hline
\end{tabular}
\end{table*}

\subsubsection{$f(R)$-Kerr black holes}

\begin{figure}[h!]
\center
\resizebox{\hsize}{!}{\includegraphics{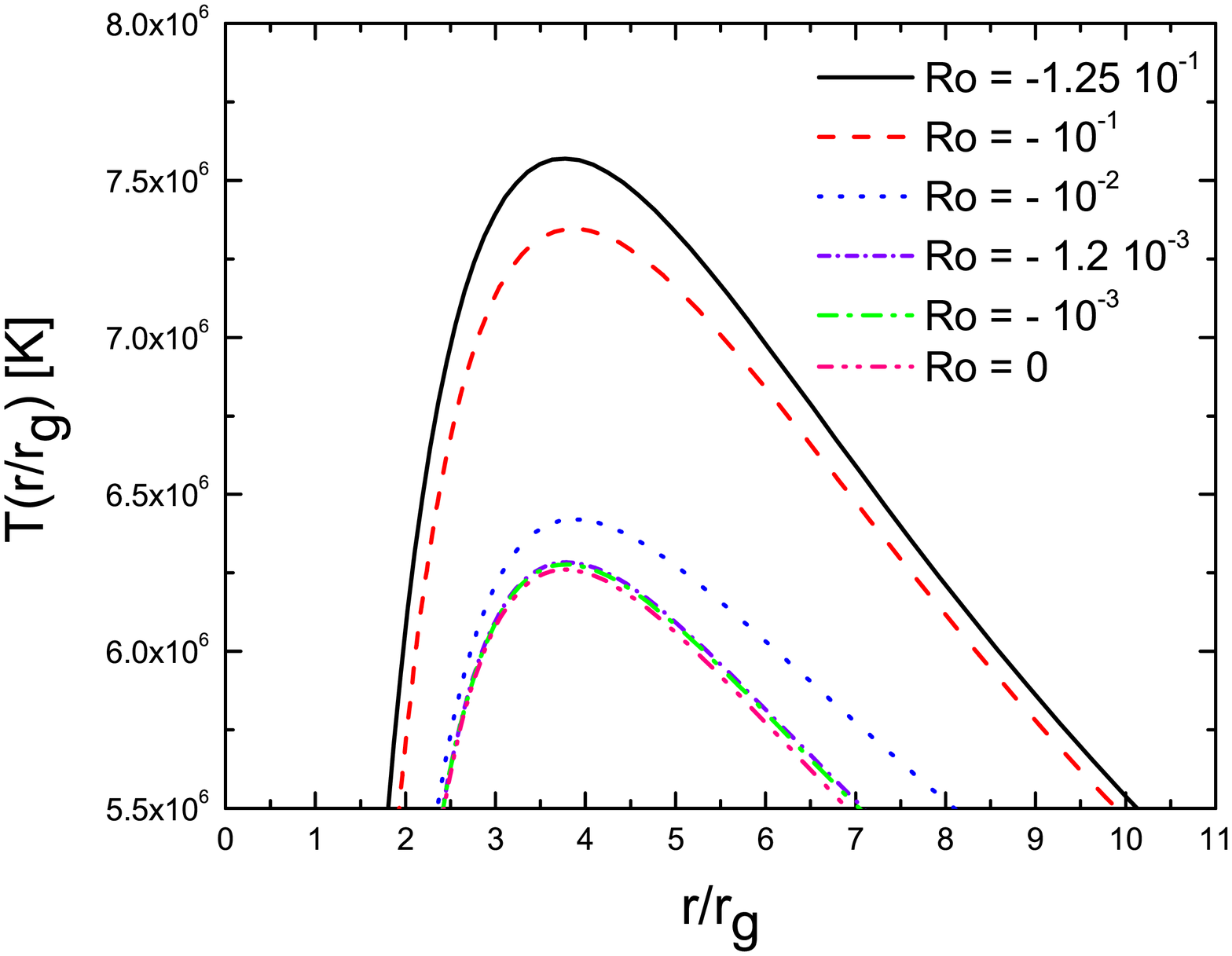}}
\caption{\label{tempkk}Plot of the temperature as a function of the radial coordinate for some values of $\mathsf{R_0}< 0 $ of a $f(R)$-Kerr black hole of angular momentum $\mathsf{a}=0.99$, corrected by gravitational redshift.}
\end{figure}

\begin{figure}[h!]
\center
\resizebox{\hsize}{!}{\includegraphics{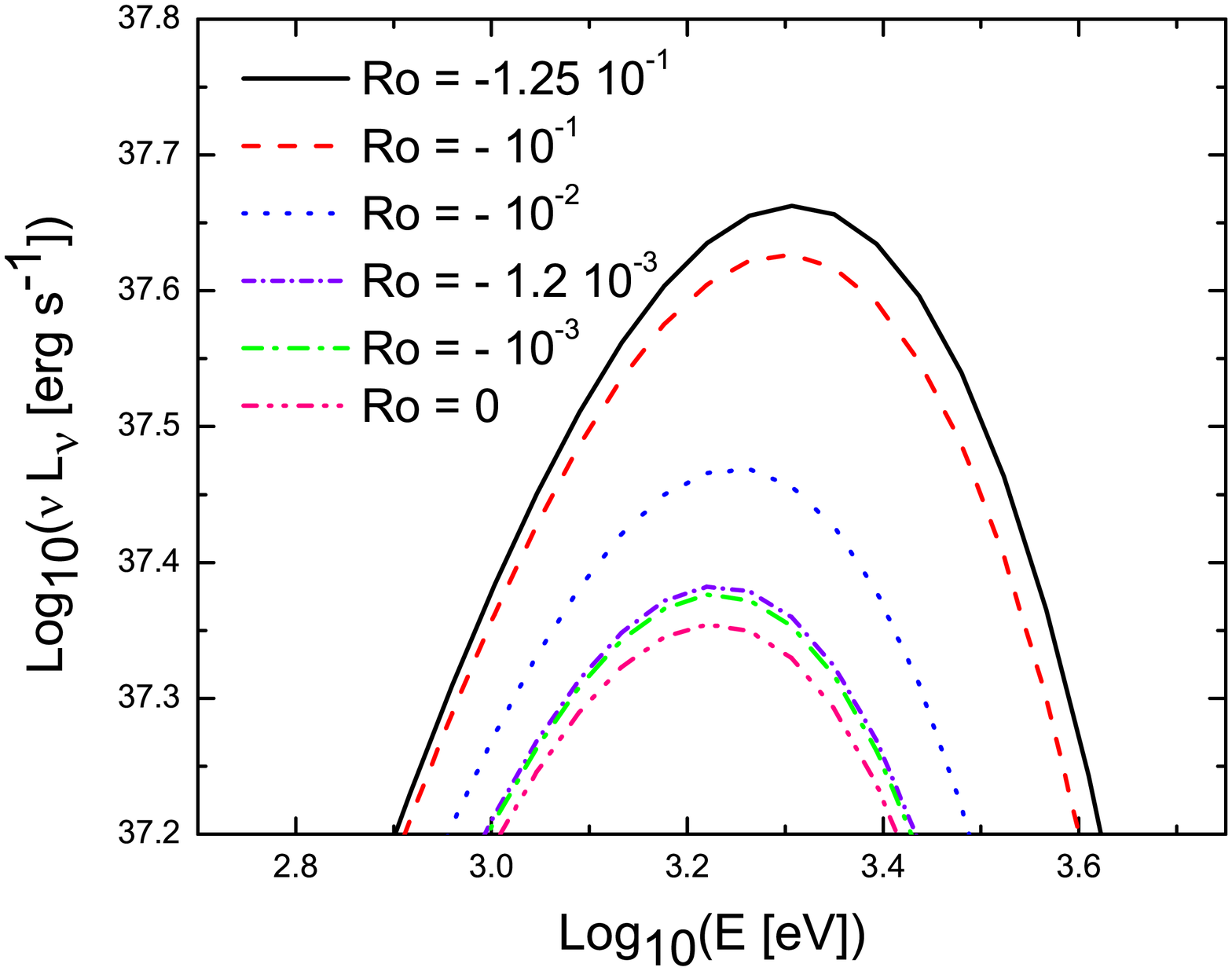}}
\caption{\label{lumikk}Plot of the luminosity as a function of the energy for some values of $\mathsf{R_0} <0$, for a $f(R)$-Kerr black hole of angular momentum $\mathsf{a}=0.99$.}
\end{figure}

We calculate next the energy flux of an accretion disk around a $f(R)$-Kerr black hole:
\begin{equation}
Q(\mathsf{x}) = - \frac{\dot{M_{0}}}{4\pi \sqrt{-g}}\frac{\Omega,_{\mathsf{x}}}{\left(\widetilde{E}-\Omega\widetilde{L}\right)^{2}}
\int^{\mathsf{x}}_{\mathsf{x}_{{\rm isco}}} \left(\widetilde{E}-\Omega\widetilde{L}\right) \widetilde{L}_{,\mathsf{x}} d\mathsf{x},
\label{kef}
\end{equation}
where:
\begin{equation}
\Omega_{,{\mathsf x}} = -36 \sqrt{3} \frac{c}{\mu^2}\eta ,
\end{equation}
\begin{equation*}
\eta \equiv \frac{\mathsf{x}^{1/2}\left\{12\mathsf{x}^3+\mathsf{a}\left[12\mathsf{a}-\mathsf{a}\mathsf{R_0}
\mathsf{x}^3-4\sqrt{36\mathsf{x}^3-3\mathsf{x}^6\mathsf{R_0}}\right]\right\}}{\left(12
\mathsf{x}^3+\mathsf{a}^2\mathsf{R_0}\mathsf{x}^3-12\mathsf{a}^2\right)^2\sqrt{-\mathsf{R_0}
\mathsf{x}^3+12}},
\end{equation*}
\begin{equation}
\widetilde{L}=\frac{2 \mu\left(l_{1}+l_{2}\right)}{\left(12+\mathsf{a}^2 
\mathsf{R_0}\right)\mathsf{x}
\sqrt{{l_{3}+l_{4}}}},
\end{equation}
\begin{equation}
\widetilde{L}_{,{\mathsf x}}=-\frac{4 \mathsf{x}\left[12\mathsf{x}^3+\mathsf{a}^2\left(-12+\mathsf{R_0}\mathsf{x}^3\right)\right]
\left( l_{5}+l_{6}+l_{7}+l_{8}+l_{9}\right)}{\left(12+\mathsf{a}^2 \mathsf{R_0}\right) \sqrt{12 \mathsf{x}^3-\mathsf{R_0} \mathsf{x}^6} \left(l_{10}+l_{11}\right)^{3/2}},
\end{equation}
\begin{equation}
\left(\widetilde{E}-\Omega\widetilde{L}\right) = \frac{12 c \sqrt{l_{3}+l_{4}}}{\left(12+\mathsf{a}^2 \mathsf{R_0}\right)
\left[12 \mathsf{x}^3+\mathsf{a}^2 \left(-12 + \mathsf{x}^3 \mathsf{R_0}\right)\right]},
\end{equation}
and
\begin{eqnarray*}
l_{1} & = & -72 \mathsf{a}^3 \mathsf{x}-216 \mathsf{a} \mathsf{x}^3+12 \mathsf{x}^3 \sqrt{36 \mathsf{x}^3-3 \mathsf{x}^6 \mathsf{R_0}},\\ 
l_{2} & = & \mathsf{a}^{2} \sqrt{36 \mathsf{x}^3-3 \mathsf{x}^6 \mathsf{R_0}}\:\left[\mathsf{a}^{2}\mathsf{x} \mathsf{R_0}+24+\mathsf{x}\left(12+\mathsf{x}^{2}\mathsf{R_0}\right)\right],\\ 
l_{3} & = & -432 \mathsf{a}^{2} \mathsf{x}^2+ \mathsf{x}^6 \left(12+\mathsf{a}^{2} \mathsf{R_0}\right)^2+48 \mathsf{a}^{3}  \sqrt{36 \mathsf{x}^3-3  \mathsf{x}^6 \mathsf{R_0}},\\ 
l_{4} & = & 144 \mathsf{a}  \mathsf{x}^2 \sqrt{36 \mathsf{x}^3-3 \mathsf{x}^6 \mathsf{R_0}}\\
&-&12 \mathsf{x}^3 \left[36 \mathsf{x}^2+\mathsf{a}^4 \mathsf{R_0}+\mathsf{a}^2 \left(36-3 \mathsf{x}^2 \mathsf{R_0}\right)\right],\\
l_{5} & = & \left[108\mathsf{a}\mathsf{x}^2\left(-24-12\mathsf{x}+5\mathsf{R_0}\mathsf{x}^3\right)
\right]\sqrt{12\mathsf{x}^3-\mathsf{R_0}\mathsf{x}^6},\\
l_{6} & = & 36\mathsf{a}^3\left(-48-36\mathsf{x}+7\mathsf{R_0}\mathsf{x}^3\right)\sqrt{12\mathsf{x}^3-
\mathsf{R_0}\mathsf{x}^6},\\
l_{7} & = & \sqrt{3}\mathsf{a}^4\mathsf{x}\left(-12+\mathsf{R_0}\mathsf{x}^3\right)
\left[-108+\mathsf{R_0}\mathsf{x}^2\left(-3+\mathsf{R_0}\mathsf{x}^3\right)\right], \\
l_{8} & = & 36 \sqrt{3} \mathsf{x}^5 \left\{72+\mathsf{x}\left[-12+\mathsf{R_0}\mathsf{x}^2\left(-15+4\mathsf{x}
\right)\right]\right\},\\
l_{9} & = & 3 \sqrt{3} \mathsf{a}^2 \mathsf{x}^2 \left\{864+\mathsf{x} l_{9a}\right\},\\
l_{9a}& = & \left[2160+\mathsf{x} l_{9a1}\right],\\
l_{9a1} & = & \left(432+\mathsf{R_0}^2\mathsf{x}^4\left(15+8\mathsf{x}
\right)-12\mathsf{R_0}\mathsf{x}\left(21+26\mathsf{x}\right)\right),\\
l_{10} & = & 144\left(-3+\mathsf{x}\right)\mathsf{x}^5+\mathsf{a}^4\mathsf{R_0}\mathsf{x}^3
\left(-12+\mathsf{R_0}\mathsf{x}^3\right)\\
&+&48\mathsf{a}^3\sqrt{36\mathsf{x}^3-3\mathsf{R_0}\mathsf{x}^6},\\
l_{11} & = & 144\mathsf{a}\mathsf{x}^2\sqrt{36\mathsf{x}^3-3\mathsf{R_0}\mathsf{x}^6}+
12\mathsf{a}^2\mathsf{x}^2l_{11a},\\
l_{11a} & = & -36+\mathsf{x}\left[-36+\mathsf{R_0}\mathsf{x}^2\left(3+2\mathsf{x}\right)\right].
\end{eqnarray*}
If we adopt for the radius of the inner edge of the disk $1.4545r_{\rm g}$, according to Eq. (\ref{re}), $r_{\rm out} \approx 16r_{\rm g}$.
From Eq. (\ref{kef})
we numerically calculate
the temperature and luminosity for the
values shown in Table \ref{tabla2}, taking into account the corrections coming from the gravitational redshift. The results are displayed in Figures \ref{tempkk}, \ref{lumikk}, and Table \ref{tabla44}.
We see that the temperature of the disk increases for smaller values of $\mathsf{R_{0}}$. The ratio of the maximum temperature between the GR and $f(R)$
cases, with $\mathsf{R_{0}} = -1.25\times 10^{-1}$, is 1.20. The peak of the emission rises a factor of 2, and the corresponding energy
is shifted towards higher energies.

\begin{figure}[h!]
\center
\resizebox{\hsize}{!}{\includegraphics{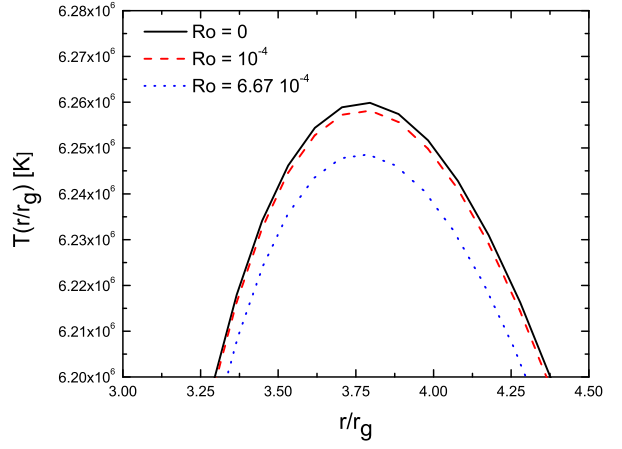}}
\caption{\label{temp-kp1}Plot of the temperature as a function of the radial coordinate  for some values of $\mathsf{R_0}>0 $, for a $f(R)$-Kerr black hole of angular momentum $\mathsf{a}=0.99$.}
\end{figure}

\begin{figure}[h!]
\center
\resizebox{\hsize}{!}{\includegraphics{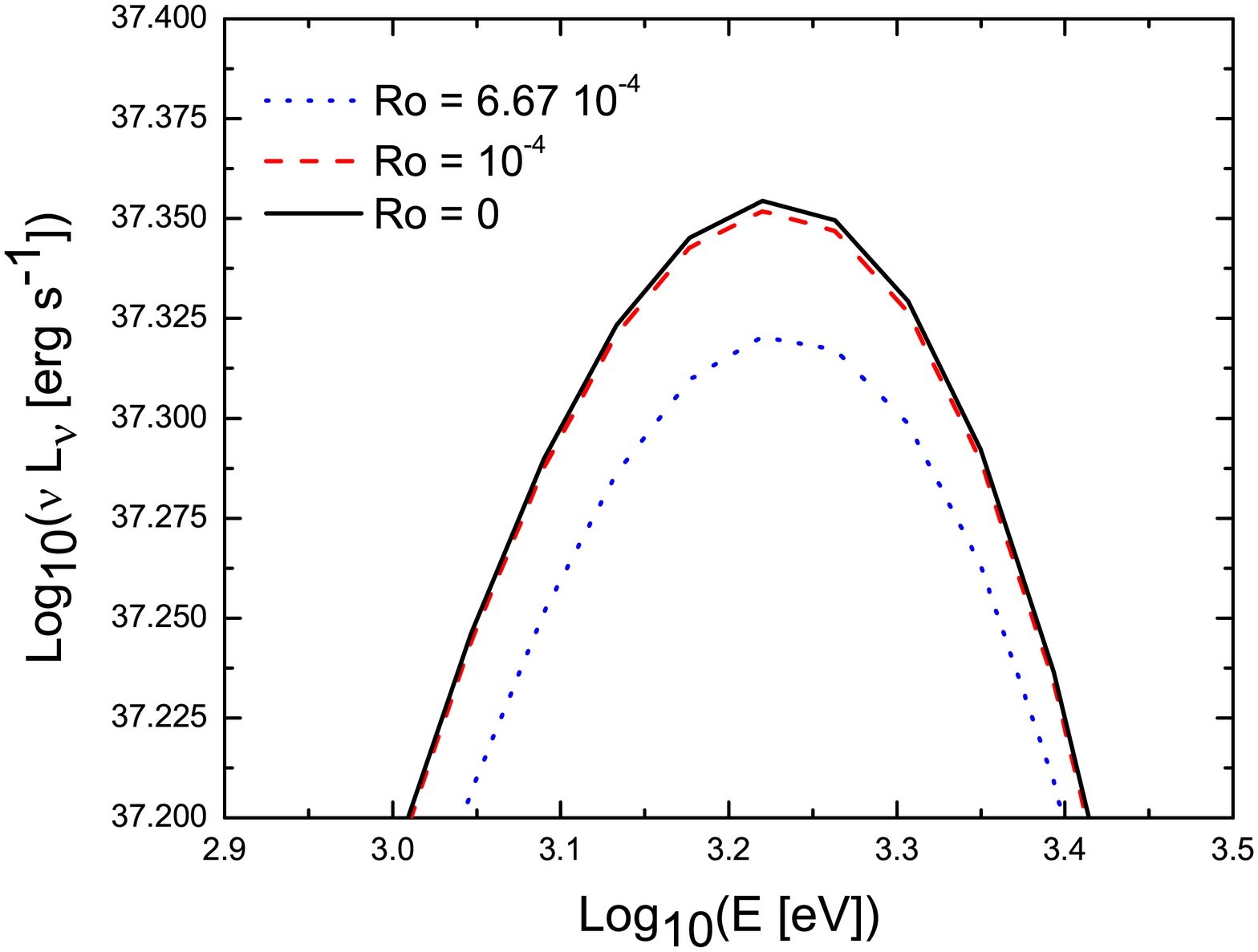}}
\caption{\label{lumi-kp1}Plot of the luminosity as a function of the energy for some values of $\mathsf{R_0} >0$, for a $f(R)$-Kerr black hole of angular momentum $\mathsf{a}=0.99$.}
\end{figure}

Since the radius of the outer edge of the disk is $16r_{\rm g}$, we see from Table \ref{tab4} that accretion disks are only possible for $\mathsf{R_0}=10^{-6}$ and $\mathsf{R_{0}} = 10^{-4}$, until up $\mathsf{R_0} = 6.67\times10^{-4}$. For such values of $\mathsf{R_0}$, we show in Table \ref{tablak} the values of the location of the last stable circular orbit, maximum temperature, luminosity, and the energy of the peak of the emission, and in Figures \ref{temp-kp1} and \ref{lumi-kp1}, the temperature and luminosity distributions respectively. As in $f(R)$-Schwarzschild black holes with positive Ricci scalar, these differences are minor. 


\begin{table*}
\caption{Location of the last stable circular orbit and maximum temperature, maximum temperature, luminosity, and energy of the peak of the emission for an accretion disk around a $f(R)$-Kerr black hole with $\mathsf{R_0}<0$ and $\mathsf{a}=0.99$.}
\centering
\label{tabla44}
\begin{tabular}{c c c c c}
\hline \hline
\rm $f(R)$-Kerr  &  $\mathsf{R_0}=0$ & $\mathsf{R_0}=-10^{-3}$ & $\mathsf{R_0}=-1.2\times10^{-3}$ \\ 
\hline
$r_{\rm isco}/r_{{\rm g}}$ & $1.4545$ & $1.4523$  &  $1.4518$\\
$r_{{\rm {\tiny Tmax}}}/r_{{\rm g}}$ & $3.79$ & $3.79$ &  $3.79$\\ 
$T_{{\rm max}}$ & $0.539\:{\rm keV}$ & $0.54119\:{\rm keV}$ & $0.54148\:{\rm keV}$\\ 
$E_{{\rm max}}$ & $1659.4\:{\rm eV}$ & $1659.4\:{\rm eV}$ & $1659.4\:{\rm eV}$\\ 
$L(E_{{\rm max}})$ & $2.26\times10^{37}{\rm erg\:s}^{-1}$ & $2.38\times10^{37}{\rm erg\:s}^{-1}$ & $2.41\times10^{37}{\rm erg\:s}^{-1}$\\ 
\hline\hline
\rm $f(R)$-Kerr  &  $\mathsf{R_0}=-10^{-2}$ & $\mathsf{R_0}=-10^{-1}$ & $\mathsf{R_0}=-1.25\times10^{-1}$ \\ 
\hline
$r_{\rm isco}/r_{{\rm g}}$  & $1.4325$ & $1.2017$ & $1.0419$\\
$r_{{\rm {\tiny Tmax}}}/r_{{\rm g}}$ & $3.85$ & $3.85$ & $3.78$ \\ 
$T_{{\rm max}}$ &  $0.553\:{\rm keV}$ & $0.663\:{\rm keV}$ & $0.652\:{\rm keV}$\\ 
$E_{{\rm max}}$  & $1833.52\:{\rm eV}$ & $2025.9\:{\rm eV}$ & $2025.9\:{\rm eV}$ \\ 
$L(E_{{\rm max}})$ & $2.94\times10^{37}{\rm erg\:s}^{-1}$ & $4.23\times10^{37}{\rm erg\:s}^{-1}$ & $4.60\times10^{37}{\rm erg\:s}^{-1}$\\ 
\hline 
\end{tabular}
\end{table*}

\begin{table*}
\caption{Values of the location of the last stable circular orbit, location in the radial coordinate of the maximum temperature, maximum temperature and luminosity, and the energy of the peak of the emission for an accretion disk around a $f(R)$-Kerr black hole with $\mathsf{R_0}>0$ and $\mathsf{a}=0.99$.}
\label{tablak}
\centering
\begin{tabular}{c c c c}
\hline\hline
\rm $f(R)$-Kerr  &  $\mathsf{R_0}=0$  & $\mathsf{R_0}=10^{-4}$ & $\mathsf{R_0}=6.67\times10^{-4}$\\
\hline
$r_{{\rm isco}}/r_{{\rm g}}$ & $1.4545$  & $1.4547$ & $1.4559$\\ 
$r_{{\rm {\tiny Tmax}}}/r_{{\rm g}}$ & $3.79$  & $3.79$ & $3.79$\\ 
$T_{{\rm max}}$ & $0.53942\:{\rm keV}$ & $0.53927\:{\rm keV}$ & $0.53843\:{\rm keV}$\\ 
$E_{{\rm max}}$ & $1659.4\:{\rm eV}$  & $1659.4\:{\rm eV}$  & $1659.4\:{\rm eV}$\\ 
$L(E_{{\rm max}})$ & $2.26\times10^{37}\:{\rm erg\:s}^{-1}$  & $2.25\times10^{37}\:{\rm erg\:s}^{-1}$ & $2.09\times10^{37}\:{\rm erg\:s}^{-1}$\\ 
\hline
\end{tabular}
\end{table*}

We proceed now to examine 
some specific forms of the function $f$, and the constraints imposed on them 
by the previous analysis. 

\section{Limits on specific prescriptions for $f(R)$}
\label{spec}

As discussed in Sections \ref{orbits-sch}
and \ref{equat}, the 
existence of Page-Thorne disks around $f(R)$ black holes
imposes the following limits on $R_0$: 
\begin{itemize}
\item $f(R)$-Schwarzschild space-time:\\
\begin{equation}\label{cond3}
\mathsf{R}_{0} \in (-\infty ;10^{-6}],
\end{equation}
\item $f(R)$-Kerr space-time:\\
\begin{equation}\label{cond4}
\mathsf{R}_{0} \in [-1.2\times10^{-3} ; 6.67\times10^{-4}].
\end{equation}
\end{itemize}

As we shall see in Sect. \ref{Discussion}, contemporary observations of Cygnus X-1 rule out accretion disks around $f(R)$-Schwarzschild  black holes, since the maximum temperature obtained in such models is lower than the inferred through observations (Gou et al. 2011). Hence, we will only consider the values of $\mathsf{R_0}$ given by expression (\ref{cond4}). 
We shall show in this section how these values lead to limits 
on the parameters of two examples of $f(R)$ theories via Eq. (\ref{condi}). We shall also impose the following viability conditions, to be satisfied by any $f(R)$ (Cembranos et al. 2011): 
\begin{equation}
-1<f'(R_{0})<0,
\label{c1}
\end{equation}
\begin{equation}
f''(R_{0})>0.
\label{c2}
\end{equation}



\subsection{$f(R) =\alpha R^{\beta}$}

The parameters $\alpha$, $\beta$ and the Ricci scalar are related by Eq. (\ref{condi}) as follows:
\begin{equation}
R_0 =  \left[\frac{1}{\alpha \left(\beta-2\right)}\right]^{\frac{1}{\beta-1}}.
\end{equation}
Introducing the adimensional parameter $\alpha ' = R_g^{\beta -1}\alpha$ with $R_g\equiv r_g^{-2}$, this equation can
be written as
\begin{equation}
\mathsf{R}_0 =  \left[\frac{1}{\alpha ' \left(\beta-2\right)}\right]^{\frac{1}{\beta-1}}.
\label{86}
\end{equation}
Notice the condition $\beta>0$ to ensure the GR limit for small values of the Ricci scalar $R$.
Let us consider first the case of a positive Ricci scalar, which leads to:\\

\textbf{Case I}
\begin{equation}\label{parte1}
\alpha '>0\:\wedge\:\beta>2,
\end{equation}
or
\begin{equation}\label{parte2}
\alpha '<0\:\wedge\:\beta<2,
\end{equation}

\begin{figure}

\includegraphics[width=8cm]{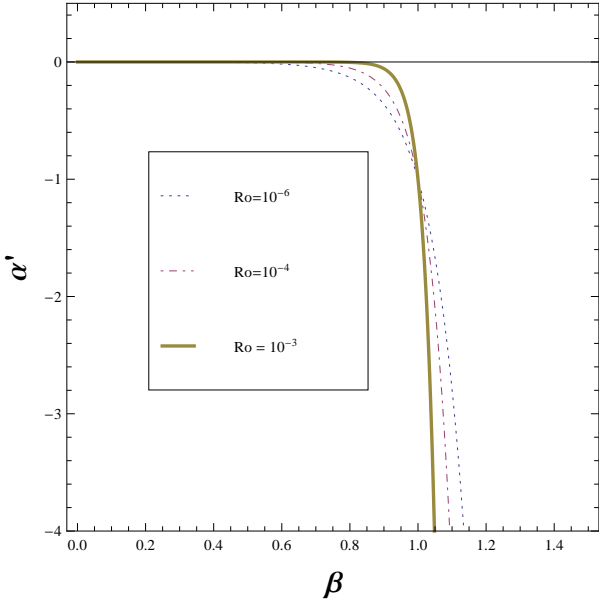}
\caption{Plot of $\alpha'$ as a function of $\beta$ for different values of $\mathsf{R_0}$.}
\label{f1}
\end{figure}

\textbf{Case II}
\begin{equation}
\alpha '<0\:\:\:\wedge\:\:\:\beta>2,
\end{equation}
or
\begin{equation}
\alpha '>0\:\:\:\wedge\:\:\:\beta<2,
\end{equation}
and $1/(\beta-1)$ an even number, that is:
\begin{equation}
\beta = 1+ \frac{1}{2n},
\end{equation}
with $n \in \mathbb{Z}$. 
By isolating $\alpha '$ from Eq. (\ref{86}), we obtain the function:
\begin{equation}\label{ab}
\alpha '(\beta) = \frac{1}{\mathsf{R_0}^{\beta-1}}\left(\frac{1}{\beta-2}\right).
\end{equation}
We show in Figure \ref{f1} the plot of $\alpha '$ as a function of $\beta$ (with $\beta >0$) for fixed values of the Ricci scalar.
We see that for $\beta \in (0,2)$, $\alpha ' \in (-\infty,0)$. For $\beta = 2$, Eq. (\ref{ab}) is not defined, while for $\beta>2$, $\alpha '$ takes large positive values. Since $\alpha$ needs to be small in order to recover GR for small values of the Ricci scalar, the case $\alpha '>0$, $\beta>2$ is discarded. Hence, we obtain the following 
restrictions on the parameters:
\begin{center}
\boxed{\alpha '\in (-\infty ;0)\:\:\:\wedge\:\:\:\beta \in (0 ;2)\:\:\:\wedge\:\:\:\mathsf{R}_0 \in (0 ;6.67\times 10^{-4}].}
\end{center}
For negative values of the Ricci scalar, from Eq. (\ref{ab}) we require that:
\begin{equation}
1-\beta = 2m,\:\:\:\Rightarrow\:\:\: \beta_{\rm odd} = 1-2m,
\end{equation}
or:
\begin{equation}
1-\beta = 2m+1,\:\:\:\Rightarrow\:\:\: \beta_{\rm even} = -2m,
\end{equation}
where $m\in \mathbb{Z}^{-}_{0}$, so that $\beta >0$. If $m = 0$, $\beta = 1$ and $\alpha ' = \alpha = -1$. These values of the parameters lead to $f(R)=-R$, which does not
reduce to GR. For $\beta = 2$, Eq. (\ref{ab}) is not defined. If $\beta \geq 3$ and is an odd number, $\alpha '$ takes positive large values, while if $\beta \geq 4$ and is an even number $\alpha '$ is large and negative. Since $\alpha '$ has to be small to recover GR for small values of the Ricci scalar, we conclude that
negative values of $\mathsf{R_0}$ are not allowed in this theory.

We now restrict the values of $\alpha$ and $\beta$ according to Eqs. (\ref{c1}) and (\ref{c2}).
The first and second derivative for the given $f$ function are:
\begin{eqnarray}
f'(R)& = &  \alpha \beta R^{\beta-1},\\
f''(R)& = & \alpha \beta \left(\beta-1\right) R^{\beta-2}.
\end{eqnarray}
The restrictions over $\alpha$ and $\beta$ that satisfy Eq. (\ref{c2}) are:
\begin{equation}\label{c21}
\alpha>0\:\:\:\wedge\:\:\: \beta>1,
\end{equation}
or
\begin{equation}\label{c22}
\alpha<0\:\:\:\wedge\:\:\:\beta\in(0,1).
\end{equation}
The condition given by Eq. (\ref{c21}) is discarded because it does not satisfy Eq. (\ref{ab}). The viability condition expressed by Eq. (\ref{c1}) takes the form:
\begin{equation}\label{nueva}
-1<\alpha\beta {R_0}^{\beta-1}<0.
\end{equation}
We can constrain the values of $\alpha$ using the latter inequality as follows:
\begin{equation*}
0< \beta <1.
\end{equation*}
By multiplying by ${R_0}^{\beta-1}\:\alpha$ the latter restrictions yields:
\begin{equation}
0 > \alpha \beta {R_0}^{\beta-2} > \alpha {R_0}^{\beta-1}.
\end{equation}
In order to satisfy Eq. (\ref{nueva}):
\begin{equation*}
\alpha {R_0}^{\beta-1} > -1\:\:\:\Rightarrow\:\:\: \alpha> \frac{-1}{{R_0}^{\beta-1}}.
\end{equation*}
If $\beta =0$, $\alpha > -\: R_0$, and for $\beta =1$, $\alpha> -1$. The set of values for $\alpha$ is $\alpha \in (-R_0,0)$.

We conclude that the values of $\alpha$ and $\beta$ that are permitted by our model as well as by the two viability conditions are:
\begin{center}
\boxed{\alpha \in (-R_0 ;0)\:\:\:\wedge\:\:\:\beta \in (0 ;1)\:\:\:\wedge\:\:\:\mathsf{R_0} \in (0 ;6.67\times 10^{-4}].}
\end{center}
 
\subsection{$f(R)= R \:\:\epsilon \ln{\frac{R}{\alpha}}$}

In this case, the parameters $\epsilon$ and $\alpha$, and the Ricci scalar are related by Eq. (\ref{condi}) in the simple form:
\begin{equation}\label{ep}
\alpha = R_0\exp{\left(\frac{1}{\epsilon}-1\right)} .
\end{equation}
Dividing by $R_g$, we obtain $\alpha '= \mathsf{R_0}\exp{\left(\frac{1}{\epsilon}-1\right)} $.
For $\mathsf{R_0}>0$, the function $\alpha '(\epsilon)$ is always positive,
while it is negative for all $\epsilon$ if $\mathsf{R_0}<0$.
The constrains over $\epsilon$ and $\alpha$ that follow from Eq. (\ref{condi}) are:
\begin{itemize}
\item $\mathsf{R_0} \in (0 ;6.67\times 10^{-4}]$\\
\begin{equation*}
\epsilon \in (-\infty ;0)\:\:\:\wedge\:\:\:\alpha ' \in (0 ;e^{-1}\;\mathsf{R_0}),
\end{equation*}
or
\begin{equation*}
\epsilon \in (0 ;\infty)\:\:\:\wedge\:\:\:\alpha '\in (e^{-1}\;\mathsf{R_0} ;\infty),
\end{equation*}

\item $\mathsf{R_0} \in [-1.2 \times 10^{-3} ;0)$\\

\begin{equation*}
\epsilon \in (-\infty ;0)\:\:\:\wedge\:\:\:\alpha '\in (-e^{-1}\;\left|\mathsf{R_0}\right|;0),
\end{equation*}
or
\begin{equation*}
\epsilon \in (0 ;\infty)\:\:\:\wedge\:\:\:\alpha ' \in (-\infty ;-e^{-1}\;\left|\mathsf{R_0}\right|),
\end{equation*}

\end{itemize}
The first and second derivative of the function $f$ take the form:
\begin{eqnarray}
f'(R) & = & \epsilon \left(1+\ln{\frac{R}{\alpha}}\right),\\ \label{f11}
f''(R) & = & \frac{\epsilon}{R}\label{f2}.
\end{eqnarray}
The condition $f''(R_0)>0$ is satisfied if $\epsilon>0\:\: \wedge R_0>0$, or $\epsilon<0\:\: \wedge R_0<0$. 
Equation (\ref{c1}) in adimensional form is:
\begin{equation}
-1<\epsilon\left(1+\ln\frac{\mathsf{R_0}}{\alpha '}\right)<0
\label{cond2}
\end{equation}
This equation, together with (\ref{f2}), yields:
\begin{itemize}
\item For $\epsilon > 0$ and $\mathsf{R_0}>0$, $\alpha ' \in \left(e\;\mathsf{R_0}  ; \mathsf{R_0}\exp\left\{\left(\frac 1 \epsilon +1\right)\right\}\right)$.

\item For $\epsilon < 0$ and $\mathsf{R_0}<0$,\\
 $\alpha ' \in \left( -e\;|\mathsf{R_0}| ; -|\mathsf{R_0}|\exp\left\{1-\frac{1}{|\epsilon|}\right\}\right)$.

\end{itemize}
Summarizing all the constraints, we have that:
\begin{eqnarray}
\mathsf{R_0} & \in &(0 ;6.67\times 10^{-4}],\;\;\;\;\epsilon >0,\;\;\;\;\alpha '\in (e^{-1}\;\mathsf{R_0} ; \infty),\\ \nonumber
\alpha ' & \in & \left(e\;\mathsf{R_0}  ; \mathsf{R_0}\exp\left\{\left(\frac 1 \epsilon +1\right)\right\}\right), \nonumber
\end{eqnarray}
and
\begin{eqnarray}
\mathsf{R_0} & \in & [-1.2\times 10^{-3};0),\;\;\;\;\epsilon <0,\;\;\;\;\alpha '\in (-e^{-1}\;|\mathsf{R_0}| ; 0), \\ \nonumber
\alpha '  & \in  &\left( - e\;|\mathsf{R_0}|; -|\mathsf{R_0}|\exp\left\{1-\frac{1}{|\epsilon|}\right\}\right).\nonumber
\end{eqnarray}
The first group of constraints is fulfilled for $\epsilon >0$, while in the second group for only $\epsilon \in (-1/2, 0)$. We conclude that for the $f(R)$ under scrutiny, the values 
\begin{equation}
\mathsf{R_0}  \in (0 ;6.67\times 10^{-4}],\;\;\;\;\epsilon >0,\;\;\;\;\alpha '\in (e \;\mathsf{R_0} ; \infty),
\end{equation}
and,
\begin{equation}
\mathsf{R_0}  \in  [-1.2\times 10^{-3};0),\;\;\;\;\epsilon \in (-1/2,0),\;\;\;\alpha ' \in  \left(-e^{-1}\;|\mathsf{R_0}|; 0\right)
\end{equation}
are allowed.

\section{Discussion}
\label{Discussion}

The results presented in Section \ref{sacc} can be compared with current observational data to derive some constraints on a given $f(R)$ theory. In order to  illustrate this assertion we shall consider Cygnus X-1, which is the most intensively studied black hole binary system in the Galaxy. A series of recent high-quality papers (Reid et al. 2011, Orosz et al. 2011, Gou et al. 2011) have provided an unprecedented set of accurate measurements of the distance, the black hole mass, spin parameter $\mathsf{a}$, and the orbital inclination of this source. This opens the possibility to constrain modified theories of gravity with rather local precision observations of astrophysical objects in the Galaxy. 


Cygnus X-1 was discovered at X-rays by Bowyer et al. (1965). Early dynamical studies of the compact object suggested the presence of an accreting black hole (e.g. Bolton 1972). The distance to Cygnus X-1 is currently estimated to be $1.86^{+ 0.12}_{- 0.11}$ kpc (Reid et al. 2011). This value  was determined via trigonometric parallax using the Very Long Baseline Array (VLBA). At this distance, the mass of the black hole is (Orosz et al. 2011) $M = 14.8 M_{\odot}$. This is the value adopted in all calculations presented in the previous sections.

The source has been observed in both a low-hard state, dominated by the emission of a hot corona (e.g., Dove et al.
1997; Gierlinski et al. 1997; Poutanen 1998), and a high-soft state, dominated by the accretion disk, which in this state goes all the way down to the last stable orbit. In the low-hard state, where the source spends most of the time, a steady, non-thermal jet is observed (Stirling et al. 2001). The jet is absent in the thermal state. Therefore, in this latter state a clearer X-ray spectrum can be obtained.     

The accretion rate and the spin parameter of the hole are $\sim 0.472\times 10^{19}$ g s$^{-1}$ and 0.99, respectively, according to estimates from a Kerr plus black-body disk model (Gou et al. 2011). These GR models yield a spectral energy distribution with a maximum at $E_{\max}\sim 1.6$ keV. On the contrary, $f(R)$-models with negative curvature correspond to a low maximum temperature, lower even than what is expected for the (unrealistic) case of a Schwarzschild black hole. Therefore, we can presume that a fit of f(R)-Kerr models to the data would also prefer high values of maximum temperature, i.e., ones with non-negative curvature. Models with accretion rates and spin close to those obtained by Gou et al. (2011) and small positive curvature seem viable, something that is consistent with an asymptotic behaviour corresponding to a de Sitter space-time endowed with a small and positive value of the cosmological constant.

Deep X-ray studies with {\it Chandra} satellite might impose more restrictive limits, especially if independent constraints onto the accretion rate become available.    


\section{Conclusions}

We have studied stable circular orbits and relativistic accretion disks around Schwarzschild and Kerr black holes in $f(R)$ gravity with constant Ricci curvature in the strong regime. We have found that stable disks can be formed only for curvatures in the ranges of $(-\infty,10^{-6}]$ and $[-1.2\times 10^{-3},6.67\times 10^{-4}]$ in the cases of Schwarzschild and Kerr black holes, respectively. Current observations of Cygnus X-1 in the soft state rule out curvature values below $-1.2\times 10^{-3}$. Additional constrains can be imposed on specific prescriptions of $f(R)$ gravity. In particular, logaritmic-gravity prescriptions are strongly constrained by observational data. Future high-precision determination of the parameters of other black hole candidates can be used to impose more restrictive limits to extended theories of gravity.


\begin{acknowledgements}
We are grateful to an anonymous referee for useful suggestions. BH astrophysics with G.E. Romero is supported by grant PIP 2010/0078 (CONICET). Additional funds comes from Ministerio de Educaci\'on y Ciencia (Spain) trhough grant AYA 2010-21782-C03-01. SEPB acknowledges support from UERJ, FAPERJ, and ICRANet-Pescara. We thank Gabriela Vila and Florencia Vieyro for comments on accretion disks.
      
\end{acknowledgements}

\end{document}